\documentclass[a4paper]{iopart}
\pdfoutput=1
\usepackage[table]{xcolor}
\usepackage{textcomp}
\usepackage{amsmath}
\usepackage{amssymb}
\usepackage{}
\usepackage{graphicx} 
\usepackage{}
\usepackage{pgf}
\usepackage{hyperref}
\usepackage{epstopdf}

\newcommand{\ket}[1]{\left|#1\right\rangle}
\newcommand{\bra}[1]{\left\langle #1\right|}
\newcommand{\ud}{\mathrm{d}}

\newcommand{\mean}[1]{\left\langle #1\right\rangle}

\newcommand{\nep}{\textrm{e}}

\def\cO{\hat{\mathcal{O}}}
\def\tr{\operatorname{Tr}}
\def\a{\alpha}
\def\bra#1{\langle #1 |}
\def\ket#1{| #1 \rangle}
\def\l#1{\lambda_{#1}}
\def \b {\beta}

\begin{document}

\title{Multipartite entanglement after a quantum quench}

\author{Silvia Pappalardi$^{1,2,3}$, Angelo Russomanno$^{4,1}$, Alessandro Silva$^{3}$ and Rosario Fazio$^{1,4}$}
\address{$^1$ Abdus Salam ICTP, Strada Costiera 11, I-34151 Trieste, Italy}
\address{$^2$ La Sapienza Universit\`a di Roma, Piazzale Aldo Moro, 5, 00185 Roma, Italy}
\address{$^3$ SISSA, Via Bonomea 265, I-34135 Trieste, Italy}
\address{$^4$ NEST, Scuola Normale Superiore \& Istituto Nanoscienze-CNR, I-56126 Pisa, Italy}

%\affiliation{Abdus Salam ICTP, Strada Costiera 11, I-34151 Trieste, Italy}
%\affiliation{La Sapienza Universit\`a di Roma, Roma, Italy}
%\affiliation{SISSA, Via Bonomea 265, I-34135 Trieste, Italy}

%\author{Angelo Russomanno}
%\affiliation{Abdus Salam ICTP, Strada Costiera 11, I-34151 Trieste, Italy}
%\affiliation{NEST, Scuola Normale Superiore \& Istituto Nanoscienze-CNR, I-56126 Pisa, Italy}

%\author{Alessandro Silva}
%\affiliation{SISSA, Via Bonomea 265, I-34135 Trieste, Italy}

%\author{Rosario Fazio}
%\affiliation{Abdus Salam ICTP, Strada Costiera 11, I-34151 Trieste, Italy}
%\affiliation{NEST, Scuola Normale Superiore \& Istituto Nanoscienze-CNR, I-56126 Pisa, Italy}

\begin{abstract}
We study the multipartite entanglement of a quantum many-body system undergoing a quantum quench. We quantify multipartite entanglement through
the quantum Fisher information (QFI) density and we are able to express it after a quench in terms of a generalized response function. For 
pure state initial conditions and in the thermodynamic limit, we can express the QFI as the fluctuations of an observable computed in the 
so-called diagonal ensemble. We apply the formalism to the dynamics of a quantum Ising chain, after
a quench in the transverse field. {In this model the asymptotic state is, in almost all cases, more than two-partite
entangled. Moreover, starting from the ferromagnetic phase, we find a divergence of multipartite entanglement for small quenches closely connected to a corresponding
divergence of the correlation length. }
\end{abstract}

\maketitle

%....................................................... INTRODUCTION .............................................................................%
\section{Introduction}

Over the last decade it has been established that a large body of information concerning quantum many-body systems can be extracted from the study of 
their entanglement properties~\cite{amico2008,calabrese2009,eisert2010}. At present, there are many examples of this connection spanning a wide spectrum 
of phenomena in quantum statistical mechanics and condensed matter physics. Entanglement, for example, is tightly connected to 
the topological properties of many-body systems~\cite{kitaev2006,li2008} and to the emergence of quantum phase transitions~\cite{osterloh2002}.
In particular, the entanglement entropy, which obeys an area law for  gapped systems, is known to 
acquire logarithmic corrections at criticality ~\cite{vidal2003,calabrese2004}.  

Understanding entanglement is also very useful for the description of the non-equilibrium dynamics of quantum many-body 
systems~\cite{polkovnikov2011}. For example, while the block entropy $S$ in the ground state of a one dimensional system
either saturates or grows logarithmically as a function of block length, Calabrese and Cardy~\cite{calabrese2005}  
have shown that after a quench, as a result of dephasing, $S$ obeys a volume law after a linear increase with time. 
The increase with time is slower in the presence of disorder~\cite{dechiara2006}, with a distinct logarithmic behaviour characterizing the many-body 
localised states~\cite{mbl-ent,mbl-ent1,mbl-ent2}. These considerations can be generalised to the case of linear ramps~\cite{kz-ent,kz-ent1,kz-ent2}, 
relevant for the Kibble-Zurek-type experiments, or to the case of periodic driving~\cite{pd-ent,pd-ent1,pd-ent2,pd-ent3,pd-ent4}

At present, most of the studies of entanglement in many body systems have focused on the bipartite case (either two-site or two-block as in the entanglement entropy) while 
much less is known about multipartite entanglement, i.e. entanglement between multiple, $M>2$, subsystems~\cite{toth2012, boileau2004}.
Although several important works point to the importance of this concept  in the understanding of collective behaviour of many-body systems~\cite{oliveira2006,
wei2005, biswas2014, hofmann2014,gulden2016,oliveira_pra_2006}, the overall picture is far less clear than for bipartite entanglement.  The main reason is that the quantification and classification of multipartite entanglement is fairly more complex and full of open problems (of interest also in mathematics, see e.g.~\cite{wong2001,meyer2002, facchi2010,osterloh2005,girolami1,oliveira_pra_2006}). Though a complete classification is 
still out of sight, very promising studies of multipartite entanglement have been performed  in  specific many-body systems~\cite{guhne2005,strobel2014}.
An overview of the field can be found in the review by G\"uhne and T\`oth~\cite{guhne2009}.

A very appealing quantity characterizing the degree of multipartite entanglement through a bound is the Quantum Fisher Information (QFI)~\cite{braunstein94,hyllus2012}. While the QFI was originally introduced to quantify phase parameter estimation, it has been shown~\cite{hyllus2012,toth2012} that certain types of multipartite entanglement 
could be inferred from its scaling with the system size (more precisely from the coefficient of the linear dependence on the 
size). Most importantly, the QFI has been shown to be related to a dynamical susceptibility for a system in a thermal state~\cite{hauke2016}. Since susceptibilities 
can be experimentally measured also for quite large systems, this suggests an easy way to experimentally probe multipartite entanglement and bypass existing protocols, which exponentially scale with the system size. An important consequence of this connection is that, in critical systems, the QFI inherits the scaling properties from the susceptibilities. Therefore the QFI in many-body systems  shows critical scaling at a quantum phase transition like 
any thermodynamic quantity~\cite{hauke2016}.
 
As mentioned above, while the dynamics of bipartite entanglement has received a lot of attention~\cite{calabrese2009}, 
the multipartite case has been much less investigated until now.  The connection between QFI and susceptibilities suggests the possibility 
to extend the study of multipartite entanglement not only to equilibrium situations but also to non-equilibrium.
With this goal in mind, in this paper we propose the first systematic analysis of multipartite entanglement of a quantum many-body system out-of-equilibrium. We consider an 
isolated quantum system subject to a quantum quench protocol: a parameter of the Hamiltonian is suddenly changed and the ensuing unitary 
evolution is studied. We show that in the long time limit one can relate the value of the asymptotic QFI to a generalized response function, 
thus generalizing the results obtained by Hauke {\em et al.}~\cite{hauke2016} to a non-equilibrium situation.  
The properties of the dynamics of multipartite entanglement after a quantum quench will be discussed in detail in the case of a one-dimensional Ising model in a transverse field. Depending on the nature and the type of quench, multipartite correlations may play a prominent role in the dynamics of 
the quantum chains. In particular, we will show that the structure of entanglement in the steady state depends crucially on whether the initial condition is ferromagnetic 
or paramagnetic. In the first case,
there is no limitation on the degree of multipartiteness achievable and the smaller the quench the higher the entanglement. In the second,
the degree of multipartiteness is both limited and maximal only close to the equilibrium critical point.
%$g_f \simeq 1$.

The paper is organized as follows. In Section~\ref{multiparty:sec} we briefly review the definition and quantification of multipartite entanglement
through the quantum Fisher information. Section~\ref{fisher_quench:sec} is the core of our paper:  there we discuss the behaviour of the 
QFI after a quantum quench. We show that, under very general conditions, it relaxes to an asymptotic value given by the fluctuations of an operator 
over the so-called diagonal ensemble density matrix, and we relate this quantity to a generalized response function.
For a thermal state, the expression we have obtained reduces to the one found by Hauke {\em et al.}~\cite{hauke2016} at thermal equilibrium. Equipped with this formalism,  in 
Section~\ref{results:sec} we discuss in detail the properties of the QFI density for the quantum Ising chain in transverse field, after a quench in the external 
field. We find that the system almost always shows multipartite entanglement; moreover the degree of multipartiteness diverges for infinitesimal quenches applied
to the system in the ferromagnetic phase. There is indeed no limitation on the degree of multipartiteness which can be achieved and this is strictly connected to a
divergence of the correlation length. Finally in Section~\ref{conclusion:sec} we draw our conclusions and discuss perspectives for future work.

\section{Multipartite entanglement and quantum Fisher information} 
\label{multiparty:sec}

In this section, in order to keep the presentation self-contained, we briefly review the classification of multipartite entanglement and we  discuss in 
detail how to measure it by means of the Quantum Fisher Information. In the last few years, the relation between QFI and entanglement 
has been used to establish a strong link between quantum metrology
and quantum information science; a comprehensive review on the problem can be found in Ref~\cite{toth2014quantum}.
In discussing the relation between multipartite entanglement and the QFI we follow the original works, Refs.~\cite{hyllus2012,toth2012}.

The structure of entanglement  for a partition of a system in more than two subsystems is very rich and a complete general classification and quantification 
of multipartite entanglement is not yet available. Therefore, in order to use a definition of multipartite entanglement as precise as possible let us follow the approach 
of Ref.~\cite{guhne2005} and,  considering a state of $N$ particles, start with the following definitions of  $k$-producible pure 
states and  $k$-particle entangled pure states.

\noindent
{\em Definition: $k$-producible pure states -}
A pure state $\ket{\psi_{k-\text{prod}}}$ is $k$ producible (producible by $k$-partite entanglement) if it can be written as $\ket{\psi_{k-\text{prod}}}
=\bigotimes_l^M\ket{\phi_l}$,  where $\ket{\phi_l}$ are non-producible states of $N_l\leq k$ particles (such that $\sum_{l=1}^MN_l=N$ and there is at least one $N_l=k$). For example, for $N=3$ particles, 1 producible, 
2 producible and 3 producible states can be constructed as
$\ket{\psi_{1-\text{prod}}} = \ket{\phi}_1\otimes\ket{\psi}_2\otimes \ket{\chi}_3$, 
$\ket{\psi_{2-\text{prod}}} = \ket{\phi}_{12}\otimes \ket{\chi}_3$, and 
$\ket{\psi_{3-\text{prod}}} = \ket{\phi}_{123}$ respectively.
 
\noindent
{\em Definition:  genuine $k$-partite entangled pure states.}
A pure state $\ket{\psi_{k-\text{ent}}}$ is called \emph{$k$-partite entangled} if it is $k$-producible, but not $(k-1)$-producible. Therefore, 
a $k$-partite entangled state can be written as a product $\ket{\psi_{k-\text{ent}}}=\bigotimes_l^M\ket{\phi_l}$,  which contains at least one state 
$\ket{\phi_l}$ of $N_l=k$ particles which {\em does not factorise}. 
In the previous examples, $\ket{\psi_{2-\text{prod}}} = \ket{\phi}_{12}\otimes \ket{\chi}_3$ is a $2-$entangled states if $\ket{\phi}_{12} 
\neq\ket{\phi_1}\otimes\ket{\phi_2}$. Similarly a Greenbergen-Horne-Zeilinger state is a $3$-entangled state. These definitions are extended to mixed 
states by the convex combination exactly as it is done for the case of bipartite entanglement \cite{horodecki2009}.  

Several different possible ways to quantify multipartite entanglement have been proposed in the literature. They include geometric measures~\cite{wei2005,
sen2010}, global entanglement measures~\cite{meyer2002,osterloh2005,oliveira2006} and there are also proposals based on concurrence~\cite{wong2001}, on the distribution of 
purities~\cite{facchi2010} for different bi-partitions or on the construction of appropriate monotones~\cite{facchi2008maximally}. 
As already mentioned, in this work we choose to focus on a specific quantity, the Quantum Fisher Information (QFI), %(originally introduced in quantum phase estimation theory) 
%\begin{equation}\label{fundamental}
%k\leq f_Q\leq k+1.
%\end{equation}
%
%{\bf Perch\'e scriviamo questa formula adesso che il lettore non sa ancora niente di $f_Q$ e $F_Q$? Vorrei scrivere due parole per motivare questa apparente anomalia.}
%This equation is very important, being the foundation of all the evaluations of entanglement multipartiteness discussed in this work.which is able to  distinguish different multipartite entanglement classes through the crucial inequality Eq.~\eqref{fundamental}.
% involves the optimal Quantum Fisher Information density $f_Q$ (introduced below): given a $(k+1)$-entangled state, one can prove that 
%
%\begin{equation}\label{fundamental}
%k\leq f_Q\leq k+1.
%\end{equation}
%
%{\bf Perch\'e scriviamo questa formula adesso che il lettore non sa ancora niente di $f_Q$ e $F_Q$? Vorrei scrivere due parole per motivare questa apparente anomalia.}
%This equation is very important, being the foundation of all the evaluations of entanglement multipartiteness discussed in this work. Therefore, we hope that
%the reader will forgive us if we have presented it to him before even defining what is the Quantum Fisher Information. Nevertheless, we will not delay in satisfying his curiosity about this point.
%

Let us define and discuss in detail the QFI (we closely follow the discussion of Ref.~\cite{braunstein1994}). Its classical
counterpart, the Fisher information, is used in information theory to evaluate how precise can be the
 estimate of a parameter $\theta$, upon which a probability distribution $P(\mu|\theta)$ depends,
from measurements of the random variable $\mu$ only. %Being more precise, the Fisher Information 
It is defined as the variance of the score function
$\partial_{\theta} \log(P(\mu|\theta))$, i.e. 
\begin{eqnarray}\label{generalFisher}
F=\sum_{\mu} (\partial_\theta \log(P(\mu|\theta)))^2 P(\mu|\theta) 
\!=\! \sum_{\mu} \frac{(\partial_{\theta} P(\mu|\theta))^2}{P(\mu|\theta)}.
\end{eqnarray} 
This quantity gives a bound on the information on $\theta$ that can be obtained from an estimator $\theta(\mu)$. In particular one can easily prove
the Cramer-Rao bound, performing $m$ measurements one gets $\langle (\Delta \theta)^2 \rangle \geq 1/(mF)$: the higher $F$ the better can in principle be our estimate of $\theta$.

It is possible to apply the same logic to quantum systems. To construct a probability distribution in the quantum case we can, for example,
prepare a probe state $\hat{\rho}$, and apply to it
a unitary transformation $\hat{\mathcal{U}}(\theta)= e^{i\theta \cO}$: $\theta$ is an unknown phase shift characterizing the transformation
and $\cO$ is the operator that generates it. %~\footnote{\textcolor{red}{From a physical point of view, we can obtain this unitary transformation by coupling our quantum
%system to a {\em classical} system which is characterised by an unknown parameter $\theta$. A quite well known example is the generation of quantum coherent states 
%of the electromagnetic field by a classical current~\cite{g_K_o_book}}{(\bf Va bene questa frase?})}. 
%
The parameter $\theta$ could now be inferred by performing a measure on the shifted probe state 
$\hat{\rho}(\theta)= \hat{\mathcal{U}}^{\dagger}(\theta) \hat{\rho} \,\,  \hat{\mathcal{U}}(\theta)= e^{-i\theta \cO}\hat{\rho}\,  e^{i\theta \cO}$.
Typically the results are given by a POVM (positive operator valued measure~\cite{Nielsen_Chuang:book}) with $q$ independent elements $\{ \hat{E}_{\mu}\}$ and corresponding
outcomes $\{\mu_i\}=\{\mu_1, \dots, \mu_q\}$. %~\cite{Nielsen_Chuang:book}. 

For each outcome $\mu_i$ we can obtain an estimator $\theta_{\text{est}}(\mu_i)$; performing $m$ realizations of the measurement process we obtain
the average $\langle \theta_{\text{est}}\rangle$ of the estimator on the resulting probability distribution and its
variance $(\Delta\theta_{\text{est}})^2=\langle \theta_{\text{est}}^2 \rangle- \langle \theta_{\text{est}} \rangle ^2 $. 
%({\bf Questa frase va bene?}) % \textcolor{red}{evaluated on all possible sets of measurements} {\bf Che significa? Chiedere a Silvia}. 
Similarly to the classical case, this
variance has been proven to obey a bound~\cite{holevo-book}:  $\Delta  \theta_{\text{est}} \geq \frac 1{\sqrt{m F}}$, where $F$ is the Fisher information 
defined as in Eq.(\ref{generalFisher}) with 
%as
%
%\begin{equation}
%\label{eq:classFI}
%	F=\sum_{\mu} \frac 1{P(\mu|\theta)} \partial_{\theta}( P(\mu|\theta))^2,
%\end{equation}
% 
the probability distribution $P(\mu|\theta)=\tr(\hat{\rho}(\theta)\hat{E}_{\mu})$. 
%$F$ gives the amount of information about 
%the parameter $\theta$ contained in the measurement $\mu$. 

It is now possible to introduce a quantity $F_Q$ characterizing the usefulness of a quantum state for phase estimation given the operator $\cO$. % and \textcolor{red}{the chosen 
%final measurement} {\bf Che significa? Che fisso la POVM? 
%Allora come posso massimizzare? Chiedere a Silvia}. 
This can be done by maximizing $F$ over all possible POVMs measurements~\cite{braunstein1994}: the result is defined as the Quantum Fisher Information.
Since $F\leq F_Q (\cO)$  the bound
$\Delta  \theta_{\text{est}} \geq \frac 1{\sqrt{m F_Q (\cO)}}$ follows (quantum version of the Cramer-Rao bound). 
For pure states, the QFI is simply given by the variance of the operator that induces the phase shift,  $F_Q(\cO)=\, 4\, \langle \Delta \cO^2 \rangle$. 
If instead we consider a mixed state as an input $\hat{\rho}=\sum_{\a}p_{\a}\ket{\lambda_{\a}}\bra{\lambda_{\a}})$ (with $p_{\a}>0 \, ,\, \sum_{\a}p_{\a}=1$),  
the quantum Fisher information can be written~\cite{braunstein1994} in terms of the eigenvalues of the input state and of the 
matrix elements of the phase shift operator $\cO$ as
\begin{equation}
\label{eq:QFI}
	F_Q (\cO)=2\sum_{\a, \b} \frac{(p_{\a}-p_{\b})^2}{p_{\a}+p_{\b}}|\bra{\l{\a}}\cO\ket {\l{\b}}|^2\,.
\end{equation}
%
%In the case of a pure state ($p_l=\delta_{l,\overline{l}}$) this expression reduces to a remarkably simple formula: 
%Therefore the Fisher information is deeply related to the quantum fluctuations of the operator which induced the phase shift.  

%The quantum Fisher information is a good measure for multipartite entanglement~\cite{hyllus2012, toth2012}. It is possible to define criteria 
%which are deeply connected to phase estimation and which allow to distinguish between different multipartite entanglement classes (for the
%definition of multipartite entanglement. 
Let us now finally come to the connection between QFI and multipartite entanglement. This was thoroughly explored in
Ref.~\cite{hyllus2012}, where the authors considered a system of $N$ $\frac 12$-spins ~\cite{braunstein1994} subject to phase shift generated
by $\cO_{lin}=\frac 12 \sum_{l=1}^N{\bf n}_l\cdot \hat{\boldsymbol{\sigma}}_l$: for each $l=1,\,\ldots,\,N$, ${\bf n}_l$ is a vector on the 
Bloch sphere and $\hat{\boldsymbol{\sigma}}_l=(\hat{\sigma}_l^x,\hat{\sigma}_l^y,\hat{\sigma}_l^z)$ is the vector of the Pauli matrices 
associated to the spin $l$. The authors found an inequality relating the multipartite entanglement properties of the considered state and the QFI optimised over all 
the possible choices of $\cO_{lin}$. For $k$-producible states, the result is
$F_Q[\rho_{\rm k-prod}] \leq sk^2+r^2$, 
where $s=\left \lfloor \frac Nk\right \rfloor$ (and $\left \lfloor x\right \rfloor$ is the largest integer smaller or equal $x$) and $r=N-sk$.
%This inequality provides a criterion for the evaluation of the multipartite entanglement properties of the considered state. 
Given a 
probe state $\hat{\rho}$, if the bound is violated, then the probe state contains useful $(k+1)$-partite entanglement. 
%Since the 
%dependence of the inequality on $k$ is monotonic, the sensitivity increases with the number of particles. 
When $k$ is a divisor of N the bound further simplifies if expressed in terms of the optimal \emph{Quantum Fisher Information density}, defined as 
$f_Q\equiv \frac{F_Q}N$: if % For a $(k+1)$-entangled state it must be
\begin{equation}
%\label{eq:QFIdensity}
f_Q>k\,,
\end{equation}
%
%At the same time if the state is $(k+1)$-entangled , then $(k+1)\geq f_Q$ (otherwise the state would be $(k+2)-$ entangled and so on). 
%We consequently obtain the fundamental inequality connecting optimal QFI density: a state is $k+1$-partite entangled {if and only if}
%
%\begin{equation}\label{fundamental}
%k< f_Q\leq k+1.
%\end{equation}
the state is at least $(k+1)$-multipartite entangled.
%
%{\bf Perch\'e scriviamo questa formula adesso che il lettore non sa ancora niente di $f_Q$ e $F_Q$? Vorrei scrivere due parole per motivare questa apparente anomalia.}
In the next Sections, we are going to apply this inequality to the study of the multipartite entanglement in a system subjected to a quantum quench.

%A simple bound for the number of entangled particles $N_Q$ is given by   $k+1=N_Q \geq f_Q$. If the system has some correlation length $\xi$, then the 
%correlator $\mean{\cO_{\bf i}\cO_{\bf j}}-\mean{\cO_{\bf i}}\mean{\cO_{\bf j}}$ for the optimal $\cO_{lin}$ will be non-zero for $|{\bf i}-{\bf j}|<\xi$
%and vanishing for $|{\bf i}-{\bf j}|>\xi$ ( ${\bf i}$, ${\bf j}$ are $d$-dimensional integer valued vectors specifying the position of the site in the lattice).One  
%can then provide the estimate $ f_Q\sim\xi^d$.
%The QFI density is an increasing function of the correlation length. This is physically sound: we have seen that the QFI density is a measure
%of multipartite entanglement which evaluates quantum correlations in the system. 

\section{Quantum Fisher Information out-of-equilibrium} 
\label{fisher_quench:sec}

The purpose of this work is to study multipartite entanglement in a many-body system in non-equilibrium conditions. In particular, we
will consider the dynamics of thermally isolated quantum systems following a quantum quench, i.e. a rapid change of the system parameters.
The system is initialised in a given (possibly mixed) many-body state, and is let free to evolve in time under the action of an Hamiltonian $\hat{H}$. 
In the thermodynamic limit, local observables and correlation functions  are expected to attain a stationary value at long times. 
This eventual stationary condition is described by the diagonal ensemble, which can be obtained as the infinite-time-average of the density matrix. 
%(for non degenerate systems).

In order to characterize multipartite entanglement both in the transient and in the stationary state, we now aim at studying the QFI in such conditions. It has been shown that lower bounds on the QFI can be computed in terms of few observable quantities ~\cite{apellaniz2015optimal}. Moreover, in the special case of thermal equilibrium, the QFI can be expressed in terms of a dynamical response function~\cite{hauke2016}: it would be 
highly desirable to have a similar expression for a generic non-equilibrium situation. Below, we generalize the result of Ref.~\cite{hauke2016} 
to a many-body system subject to a quantum quench, and show that also in this case the QFI can be expressed in terms of a \it generalized \rm response function 
of the operator $\cO$ generating the phase shift. 

%In the present work, the non-equilibrium arises from the dynamical evolution following a quantum quench.  We suppose that 
In order to obtain this result, let us start by choosing  a basis for the initial state that diagonalizes the density matrix $\hat{\rho}=\sum_{\a}p_{\a} \ket{\l{\a}} \bra {\l{\a}}$. 
If the initial state is a thermal one relative to the initial Hamiltonian $\hat{H}_0$, then $\hat{H}_0 \ket{\lambda_{\alpha}}=E^0_{\alpha}\ket{\lambda_{\alpha}}$ and 
$p_{\a}=e^{-\beta E^0_{\a}}/Z$ is the standard Gibbs weight.
The state is then time evolved with the final Hamiltonian $\hat{H}$, which leads to  
\begin{equation} \label{rho_time:eqn}
  \hat{\rho}(t)=\sum_{\a}p_{\a} \ket{\l{\a}(t)}\bra {\l{\a}(t)}=\sum_{i\,j}a_{ij}e^{-i(E_i-E_j)t} \ket{\psi_i} \bra {\psi_j}\,,
\end{equation}
where $\ket{\l{\a}(t)}=\nep^{-i\hat{H} t}\ket{\l{\a}}$ and
$\{\ket{\psi_i}\}$ and $\{E_i\}$ are the eigenvectors and eigenvalues of $\hat{H}$. In particular,
$a_{ij}\equiv\sum_\alpha p_\alpha\left\langle\psi_i\right.\ket{\l{\a}}\bra {\l{\a}}\left.\psi_j\right\rangle$.
Using Eq.~(\ref{eq:QFI}) we can write the quantum Fisher information at time $t$ as
\begin{equation}
\label{eq:QFI_time}
F_Q (\cO,t)=2\sum_{\alpha\beta} \frac{(p_\alpha-p_\beta)^2}{p_\alpha+p_\beta}|\bra {\l{\a}(t)} \cO\ket{ \l{\beta}(t)}|^2\,.
\end{equation}

Focusing now on thermal initial states and using the identity $(p_{\a}-p_{\beta})/(p_{\a}+p_{\beta})=\tanh [\beta(E^0_{\beta}-E^0_{\a})/2]$ it is easy 
to show that
\begin{equation}\label{QFI:generalized}
F_Q (\cO,t)=\frac{4}{\pi} \int_0^{+\infty}d\omega\;\tanh\left[\frac{\beta\omega}{2}\right]\;\tilde{\chi}^{\prime\prime}(t,\omega)\,,
\end{equation}
where 
\begin{equation}\label{FTchi}
\tilde{\chi}^{\prime\prime}(t,\omega)=\pi\!\sum_{\a,\beta}(p_{\a}-p_{\beta})  |\bra{\lambda_{\a}(t)}\cO\ket{\lambda_{\beta}(t)}|^2 \delta(\omega+E^0_{\a}-E^0_{\beta})\,.
\nonumber
\end{equation}
In particular $ \tilde{\chi}^{\prime\prime}(t,\omega)=-{\rm Im}[  \tilde{\chi}(t,\omega)]$ where the latter is the Fourier transform
with respect to $\tau$ of the generalized retarded correlation function
\begin{eqnarray}\label{correlammerda:eqn}
\tilde{\chi}(t,\tau)=-i\theta(\tau){\rm Tr}\left[\hat{\rho}\; [\cO(t,\tau),\cO(t,0)]\right],
\end{eqnarray}
where $\cO(t,\tau)=e^{i\hat{H}_0\tau} e^{i\hat{H}t} \cO e^{-i\hat{H}t} e^{-i\hat{H}_0\tau}$.

The previous equations are one of the main results of this paper and generalize the equilibrium results obtained in Ref.~\cite{hauke2016} to the case of a quantum quench. 
Notice that in general Eq.~\eqref{correlammerda:eqn} is the linear response function at time $\tau$ of the operator $\hat{\mathcal{O}}_t\equiv\nep^{i\hat{H} t}\hat{\mathcal{O}}\nep^{-i\hat{H} t}$. In the case of thermal equilibrium ($\hat{H}_0=\hat{H}$), one straightforwardly obtains that $\tilde{\chi}(t,\tau)=-i\theta(\tau){\rm Tr}\left[\hat{\rho}\; [\cO(\tau),\cO(0)]\right]$:
the QFI comes from the imaginary part of the standard response function associated to the phase-shift operator~\cite{hauke2016}. In the equilibrium case, the validity of the fluctuation dissipation theorem is crucial. Out-of-equilibrium, where in general this theorem does not hold, the QFI cannot be written as a dynamical susceptibility, as discussed in~\ref{equilibrium:sec}. %Nevertheless, out-of-equilibrium $\tilde{\chi}''(t, \omega)$ does not represent a linear response function and actually can not be written as a dynamical susceptibility averaged on the diagonal ensemble, see Appendix \ref{equilibrium:sec}.
Moreover, whenever the initial state is a pure state, as for quenches at zero temperature, we can easily obtain from Eq.(\ref{QFI:generalized})-(\ref{FTchi}) that
\begin{eqnarray}\label{zerotQFI}
F_Q(\cO,t)=4\, \langle\, (\Delta \cO(t))^2\, \rangle.
\end{eqnarray}

Equations~(\ref{QFI:generalized}) and (\ref{zerotQFI}) allow us to study the Quantum Fisher Information both in the transient and in the stationary state attained after a quantum quench.
In order to find an explicit formula for the stationary state QFI,  let us focus on the zero temperature case.
In this case, considering the thermodynamic limit, we can show that for most systems
\begin{equation} \label{eq:QFI_relax_p1}
  F_Q (\cO,\infty)=4\tr[\cO^2\hat{\rho}_{\rm d}]- 4\tr [\cO\hat{\rho}_{\rm d}]^2=4\langle \Delta \cO^2 \rangle_{\rm d} \ ,
\end{equation}
where we have introduced the diagonal ensemble~\cite{polkovnikov2011} 
\begin{equation} \label{diagonens:eqn}
  \hat{\rho}_{d}\equiv\sum_i|\bra {\l{0}}\left.\psi_i\right\rangle|^2\ket{ \psi_i}\bra {\psi_i}\,.
\end{equation}
In order to prove this, let us start from the observation that  if $F_Q(\cO,t)$ attains a stationary value as $t \rightarrow \infty$
then this is going to be given by its time average $\overline{F_Q(\cO)}=\lim_{T\rightarrow+\infty}\;1/T\int_0^T dt F_Q(\cO,t)$. Taking explicitly the time average of Eq.~(\ref{zerotQFI}), we get
\begin{equation} \label{O_diaga:eqn}
\overline{\langle(\cO(t)-\langle \cO(t) \rangle)^2 \rangle} = \langle (\Delta \cO)^2 \rangle_d - \overline{(\Delta \langle \cO \rangle)^2}\,,
\end{equation}
where $\langle (\Delta \cO)^2 \rangle_d=\langle (\cO-\langle \cO \rangle_d)^2\rangle_d$ are the fluctuations computed with respect to the diagonal ensemble,
and $\overline{(\Delta \langle \cO \rangle)^2}=\overline{(\langle \cO(t) \rangle - \overline{\langle \cO(t) \rangle})^2}$ are the temporal fluctuations of the average.
Notice now that if the average $\langle \cO(t)\rangle$ attains a well defined stationary value at large times then $\overline{(\Delta \langle \cO \rangle)^2}=0$
and therefore the equality Eq.(\ref{eq:QFI_relax_p1}) follows. In order to figure out under which conditions this happens let us focus on 
\begin{equation}
  \langle \cO (t) \rangle=\sum_{i\,j}c_{i}c_{j}^{\,*}\bra {\psi_i} \cO\ket{ \psi_j}\nep^{-i(E_i-E_j) t}\,,
\end{equation}
where  $c_{i}=\langle \lambda_0 | \psi_i \rangle$. We can split the sum on the right hand side in two parts, a diagonal one
\begin{equation} \label{diagonal:eqn}
  \mathcal{O}_{d} = \sum_{i} |c_i|^2 \langle \psi_i |\cO| \psi_i\rangle
\end{equation}
and an off-diagonal one which can be written as
\begin{eqnarray} \label{off-diag:eqn}
  &&\mathcal{O}_{({\rm off-diag})}(t) = \int_{-\infty}^\infty F(\Omega)\nep^{-i\Omega t}\quad{\rm with}\nonumber\\ 
    &&F(\Omega)\equiv\sum_{i\neq j}c_{i}c_{j}^{\,*}\bra {\psi_i} \cO\ket{ \psi_j}\delta\left(\Omega-(E_i-E_j)\right)\,.
\end{eqnarray}
In analogy with Refs.~\cite{ziraldo2012,ziraldo2013}, we see that in the thermodynamic limit the delta
functions contained in $F(\Omega)$ can merge making of $F(\Omega)$ a smooth function. Formally this
occurs when the {point spectrum} 
of $\hat{H}$ becomes a continuum, i.e.  in a many-body context for clean systems 
in the thermodynamic limit~\cite{ziraldo2012,ziraldo2013,
russomanno2012}. Under these conditions, Riemann-Lebesgue lemma applies and 
we see that the off-diagonal contribution Eq.~\eqref{off-diag:eqn} vanishes in the limit $t\to\infty$, leading to relaxation to 
a well defined asymptotic value given by Eq.(\ref{diagonal:eqn}).

Let us conclude this section re-expressing the QFI in the stationary state in terms of 
the Keldysh component of the response function
%Given an operator $\hat{A}$, a system in a state $\hat{\rho}$ and an Hamiltonian 
%acting on the system $\hat{H}_0$, the Keldysh component of the response is given by
%
\begin{equation}
\label{eq:keld}
\chi_K(\tau,\cO)= \frac 12  \tr [ \hat{\rho}_{\rm d}\{\delta\cO(\tau), \delta\cO \}]
\end{equation}
where $\delta \cO=\cO-\langle \cO\rangle_{d} $. 
Considering the Fourier transform and using a Lehmann representation 
we find
%Our focus will be on pure state, being the mixed case much more complicated than we expected.
%
\begin{eqnarray} \label{lehman:eqn}
  \chi_K(\omega,\cO)&=&\frac{1}{2}\sum_{i,j}|c_i|^2\left| \delta \cO_{ij}\right|^2\big[\delta(\omega+E_i-E_j)\nonumber\\
   &+&\delta(\omega+E_j-E_i)\big]\,,
\end{eqnarray}
Integrating over $\omega$  it is then easy to check that
\begin{equation} \label{FQK:eqn}
  F_Q (\cO,\infty)=4\langle \Delta \cO^2 \rangle_{\rm d}=4\int_{-\infty}^\infty\chi_K(\omega,\cO)\,\ud\omega\,.
\end{equation}
%
%At equilibrium and  $T=0$ one recovers the result obtained by Hauke {\em et al.}~\cite{hauke2016} (see Appendix~\ref{equilibrium:sec}).

%%%%%%%%%%%%%%%%%%%%%%%%%%%%%%%%%%%%%%%%%%%%%%%%%%%%%%%%%%%%%%%%%%%%%%%%%%%%%%%%%%%%%%%%%%%%%%%%%%%%%%%%%%%%%%%%%%%%%%%%%%%%%%%%%%%%%%%%%%%%%%%%%%%%%%%%%%%%%%%%%%%
\section{Multipartite entanglement in the Ising model after a quantum quench}
\label{results:sec}

After having introduced the necessary formalism we now move to the discussion of multipartite entanglement  in the 
specific case of a quantum quench in a one-dimensional Ising model~\cite{Lieb_AP61,pfeuty1970} described by the Hamiltonian
\begin{equation}  \label{h11}
  \hat{H}(t) =-\, \frac{1}{2}\sum_{j=1}^{L}\left(J\hat{\sigma}_j^x\hat{\sigma}_{j+1}^x + g(t) \hat{\sigma}_j^z\right)  \;.
\end{equation}
Here $\hat{\sigma}^{x,z}_j$ are spin operators at site $j$ of a chain of length $L$ with boundary conditions which can be periodic 
(PBC) $\hat{\sigma}^{x,z}_{L+1}=\hat{\sigma}^{x,z}_1$ or open (OBC) $\hat{\sigma}^{x,z}_{L+1}=0$, and $J$ is a longitudinal coupling ($J=1$ from now on).
%Let us briefly review the equilibrium properties of the Hamiltonian (\ref{h11}) with a static and homogeneous transverse field, $g(t)=g_0$.
This model 
has two gapped phases: a ferromagnet ($\left|g_0\right|<1$) and a paramagnet ($\left|g_0\right|>1$), separated by a quantum phase transition at 
$g_c=1$. 

Thanks to the Jordan-Wigner mapping~\cite{Lieb_AP61,pfeuty1970}, this system can be shown to be equivalent to a non-interacting fermion model.
Integrability allows to easily address not only the statics, but also the dynamics after a quantum quench, as elucidated, for instance, in Ref.~\cite{calabrese2005}.
Consistently with integrability, after a quantum quench in the thermodynamic limit, all the local observables have been demonstrated to asymptotically relax~\footnote{In order to get this result it is enough to show that all the two-point fermionic correlation functions undergo such a relaxation. Being the Hamiltonian quadratic and the state Gaussian, this implies asymptotic relaxation for all the local observables.} to a condition described by a generalized Gibbs ensemble (GGE)~\cite{calabrese2011,caneva2011,gge,gge4,gge1,gge3,gge2,gge5} (i.e. the density matrix maximizing the entropy provided all the constants of motion of the integrable system are conserved). The GGE is the form acquired in this case by the diagonal ensemble we are interested in (Eq.~\eqref{diagonens:eqn}):
using the results of Refs.~\cite{calabrese2011,caneva2011}, we can explicitly use it and evaluate the asymptotic QFI through Eq.~\eqref{O_diaga:eqn} and following.

In order to estimate the multipartite entanglement~\cite{hyllus2012,toth2012}, we optimize the QFI density over operators
of the form
\begin{equation}
  \cO_{lin} = \frac 12 \sum_{l=1}^N{\bf n}_l\cdot \hat{\boldsymbol{\sigma}}_l\,,
\end{equation}
where ${\bf n}_l$ are vectors with unit norm and $\hat{\boldsymbol{\sigma}}_l=(\hat{\sigma}_l^x,\hat{\sigma}_l^y,\hat{\sigma}_l^z)$ is 
the vector of the Pauli matrices associated to the spin $l$. We focus on a translationally invariant system without antiferromagnetic order 
and our dynamics starts with uniform initial conditions, therefore in the following we will assume ${\bf n}_l={\bf n}$ and optimize over its direction.
Substituting the expression for $\cO_{lin}$ in the expression for the QFI, we get
\begin{eqnarray} \label{sommazza:eqn}
 F_Q (\cO_{lin}) = 4 \sum_{\alpha=x,y,z} (n^{\alpha})^2 \langle \Delta(\hat{S}^{\alpha})^2 \rangle
 \end{eqnarray}
where we have defined the operators 
$
  \hat{S}^\alpha\equiv\frac{1}{2}\sum_{j=1}^N\hat{\sigma}_j^\alpha 
$
 We note that in Eq.~\eqref{sommazza:eqn} terms of the form 
$n_\alpha n_\beta\left( \mean{\hat{S}^\alpha\hat{S}^\beta}-\mean{\hat{S}^\alpha}\mean{\hat{S}^\beta}\right)$ with $\alpha\neq \beta$
are always vanishing if they are not present in the initial state (the symmetry of the Hamiltonian that governs the evolution prevents the build up of 
such correlations).
%In order to get an estimate of the multipartite entanglement, we have to optimise the quantum Fisher information only over 
%the three total spin components. If we 
We finally optimize Eq.~\eqref{sommazza:eqn} over all the possible directions of the unit norm vector ${\bf n}$ \footnote{This can be done, for instance, by representing ${\bf n}$ in polar coordinates. Another possible way is to look at Eq.~\eqref{sommazza:eqn} as the expectation over an unit vector of a $3\times3$ Hermitian
matrix. This is maximized taking ${\bf n}$ as the eigenvector with maximum eigenvalue and the maximum is given by this eigenvalue.} and obtain that the optimal QFI can be written as
\begin{equation} \label{ottimazza:eqn}
  F_Q = 4\max_{\alpha=x,y,z}\left[\mean{\Delta (\hat{S}^\alpha)^2}\right]=\max_{\alpha=x,y,z}F_Q (\hat{S}^\alpha)\,. 
\end{equation}
In particular, $\hat{S}^x/L$ is
the order parameter of the ferromagnetic transition: when the thermodynamic limit is considered, in the phase $g_0<1$, its expectation
over the ground state is non-vanishing~\cite{pfeuty1970}.

Below, we restrict our attention to the behaviour of the QFI density in the diagonal ensemble; we will give later some hint about the evolution towards this stationary condition.
%As we elucidate in the following sections, for all the operators $\hat{S}^\alpha$ with $\alpha=x,y,z$,
 It is now possible to evaluate 
% using a Jordan-Wigner transformation and studying a
%free-particle model in the fermionic representation. Given the total spin operator $S^{\alpha}$, and considering that the system is 
the Quantum Fisher Information density $f_Q(\hat{S}^{\alpha},\infty)=\mean{(\Delta \hat{S}^\alpha)^2}_{d}/L$ for translationally invariant systems as
%
% \begin{align} \label{tua_sorella:eqn}
%f_Q(\hat{S}^{\alpha},t)&= \frac 4L(\, \langle \hat{S}^{\alpha}\, ^2 \rangle_t -  \langle \hat{S}^{\alpha}\rangle_t ^2\, )  = 
%\frac 1L \, \sum_{j,\,l=0}^{L-1} \, \langle \sigma_l^{\alpha} \sigma_j^{\alpha} \rangle_t - \langle \sigma_l^{\alpha}\rangle_t\langle \sigma_j^{\alpha}\rangle_t\, .  
%\end{align}
%
%
%\noindent Considering that the system is translationally invariant we find
%
\begin{eqnarray}
\label{eq:qfi_gen_ord}
%\boxed{
f_Q(\hat{S}^{\alpha},t)&=&1+  2\, \sum_{n=1}^{L-1}  G^{\alpha}_n(t)\nonumber\\
&\stackrel{\scriptscriptstyle t\to \infty} {\longrightarrow}& f_Q(\hat{S}^{\alpha},\infty)= \, 1 +2 \, \sum_{n=1}^{\infty} \, G^{\alpha}_n(\infty)\,.%}
\end{eqnarray}
where we have defined $G^{\alpha}_n(t) = \langle \sigma^{\alpha}_j\sigma^{\alpha}_{j+n}\rangle_t^c$ as the connected spin correlation function at time $t$ and we have 
and exploited the translational invariance of the problem and the inversion symmetry $G^{\alpha}_n(t)=G^{\alpha}_{-n}(t)$. The asymptotic condition is 
reached only in the thermodynamic limit (see \ref{app:thermo} for a detailed discussion) and it is given in terms of the diagonal-ensemble GGE density matrix 
Eq.~\eqref{diagonal:eqn}: $G^{\alpha}_n(\infty) = \langle \sigma^{\alpha}_j\sigma^{\alpha}_{j+n}\rangle^c_{d}$. 

%(We do not need to put the $\Delta$ symbol inside the expectation value because, out of equilbrium, $\mean{\hat{S}^\alpha}=0$ for all $\alpha=x,y,z$). 
%In order to have an estimate of the multipartite entanglement
%according to Eq.~\eqref{eq:QFIdensity},
%we will focus on the QFI  density optimized over the three directions $x,y,z$, which in this case is defined as 
%
%$$
%  f_Q(\infty)=\frac{F_Q(\infty)}{L}\,.
%$$
%

%Here we only provide
%the dependence of the maximum
%In the light of previous discussion, maximizing  over the operators, the resulting quantum Fisher information density will provide us an 
%estimate of the multipartite entanglement. We see in Fig. \ref{fig:max_QFI} the asymptotic Quantum Fisher information density for 
%different initial $g_0$. We will elucidate the details on the calculation of $f_Q(\hat{S}^\alpha,\infty)$ with $\alpha=x,y,z$ in the next section. 

\begin{figure}[htpp] 
\centering
 \includegraphics[width=13cm]{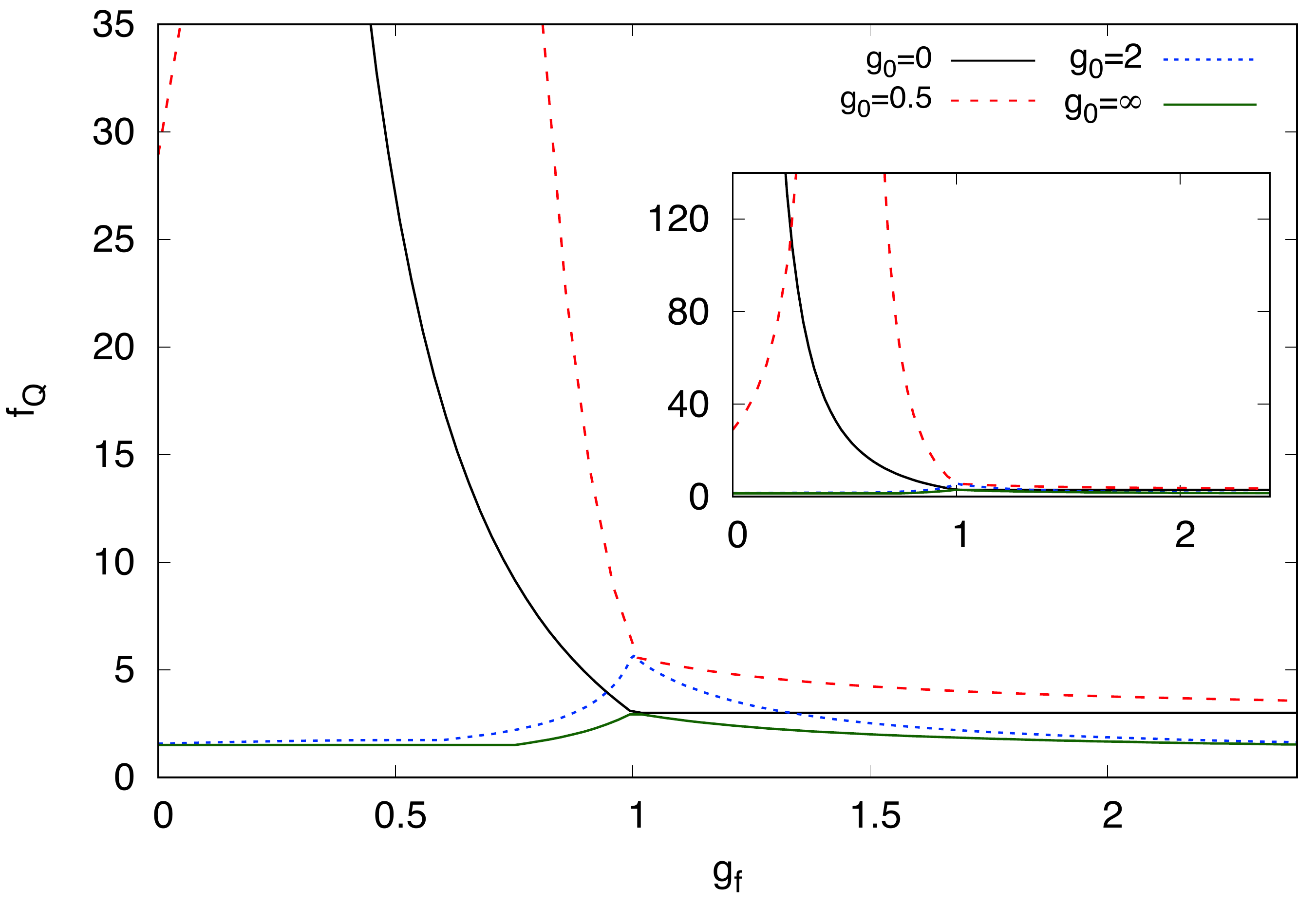}
% \label{fig:minipage1}
\caption{%{\bf Angelo: Silvia, per piacere, togli le intestazioni dalle figure (c'\`e tutto nelle didascalie). Controlla anche l'ortografia delle label.}Summary of the results for the optimal QFI asymptotic density, for different initial fields $g_0$. 
We plot the maximum over the 
three total spin components $\hat{S}^{x}, \hat{S}^{y} \, \text{and} \, \hat{S}^{x}$ from Eq.(\ref{ottimazza:eqn}) for different values of $g_0$. It is evident a strong dependence on the initial conditions: $f_Q(\infty)$ it is limited when $g_0$ is in the paramagnetic phase, whereas it diverges as $\delta g^{-2}$ for $g_0$ in the ferromagnetic one (see Sec.(\ref{sec:numer}-\ref{sec:small_quenches}) ). In the inset we plot $f_Q$ in a larger scale, in order to emphasize this divergent behavior in the ferromagnetic phase.}
\label{fig:max_QFI}
\end{figure}

%Given the total spin operator $S^{\alpha}$, and considering that the system is translationally invariant, the Quantum Fisher Information density reads :
%
% \begin{align} \label{tua_sorella:eqn}
%f_Q(\hat{S}^{\alpha},t)&= \frac 4L(\, \langle \hat{S}^{\alpha}\, ^2 \rangle_t -  \langle \hat{S}^{\alpha}\rangle_t ^2\, )  = 
%\frac 1L \, \sum_{j,\,l=0}^{L-1} \, \langle \sigma_l^{\alpha} \sigma_j^{\alpha} \rangle_t - \langle \sigma_l^{\alpha}\rangle_t\langle \sigma_j^{\alpha}\rangle_t\, .  
%\end{align}
%
%
%\noindent Considering that the system is translationally invariant we find
%
%\begin{eqnarray}
%\label{eq:qfi_gen_ord}
%\boxed{
%f_Q(\hat{S}^{\alpha},t)&=&1+  2\, \sum_{n=1}^{L-1}  G^{\alpha}_n(t)\nonumber\\
%&\stackrel{\scriptscriptstyle t\to \infty} {\longrightarrow}& f_Q(\hat{S}^{\alpha},\infty)= \, 1 +2 \, \sum_{n=1}^{\infty} \, G^{\alpha}_n(\infty)\,.%}
%\end{eqnarray}
%

It is possible to explicitly evaluate this correlator: we need to use the Jordan-Wigner transformation in order to write the $\hat{\sigma}^{\alpha}_j$ operators 
in terms of the fermionic ones; then we have to apply the Wick theorem to the resulting expectation of a string of fermionic operators on the 
Gaussian asymptotic GGE state $\hat{\rho}_{d}$~\cite{calabrese2011}. The details of the calculation can be found in 
Refs.~\cite{barouch1971,sengupta2004}. \medskip

As we will show in the rest of this Section, multipartite entanglement has a prominent role in the steady state after a quantum quench. Depending 
on the value of the final external field, it may involve a macroscopic number of spins. In Fig.\ref{fig:max_QFI} we summarize out results for the asymptotic state. We find that the attained state after a quantum quench is never separable. Moreover, 
we notice a strong dependence on the initial conditions: this is not surprising given the integrability of the system. For quenches starting from the ferromagnetic phase the QFI density diverges as $\delta g^{-2}$ for the so-called \emph{small quenches}, i.e. $\delta g= g_f -g_0 \to 0$. In the paramagnetic phase the multipartiteness is limited and maximal at the equilibrium critical point. \\
In the next sections we will present the details on the calculation of $f_Q(\hat{S}^\alpha,\infty)$ with $\alpha=x,y,z$ and comment the behaviour of the QFI density as a function of time. Furthermore we discuss the entanglement divergence in the ferromagnetic phase, through a perturbative expansion in $g_f-g_0$.
%
%......................................................... QFI density for the order parameter ...................................................................%
%
\subsection{Exact results for the asymptotic QFI for $g_0=0$ and $g_0=\infty$}\label{sec:exa}

In the two limiting cases of quenches starting from ${g_0=\infty}$ (fully polarised initial state) or ${g_0=0}$ (maximally ordered classical point),  the asymptotic correlation function 
after the quench  can be expressed in very nice and simple
forms (see Ref.~\cite{suzuki-book} for the details of the calculation) leading, for $g_0 = \infty$, to
\begin{figure}[htpp]
\centering
 \includegraphics[width=10.5cm]{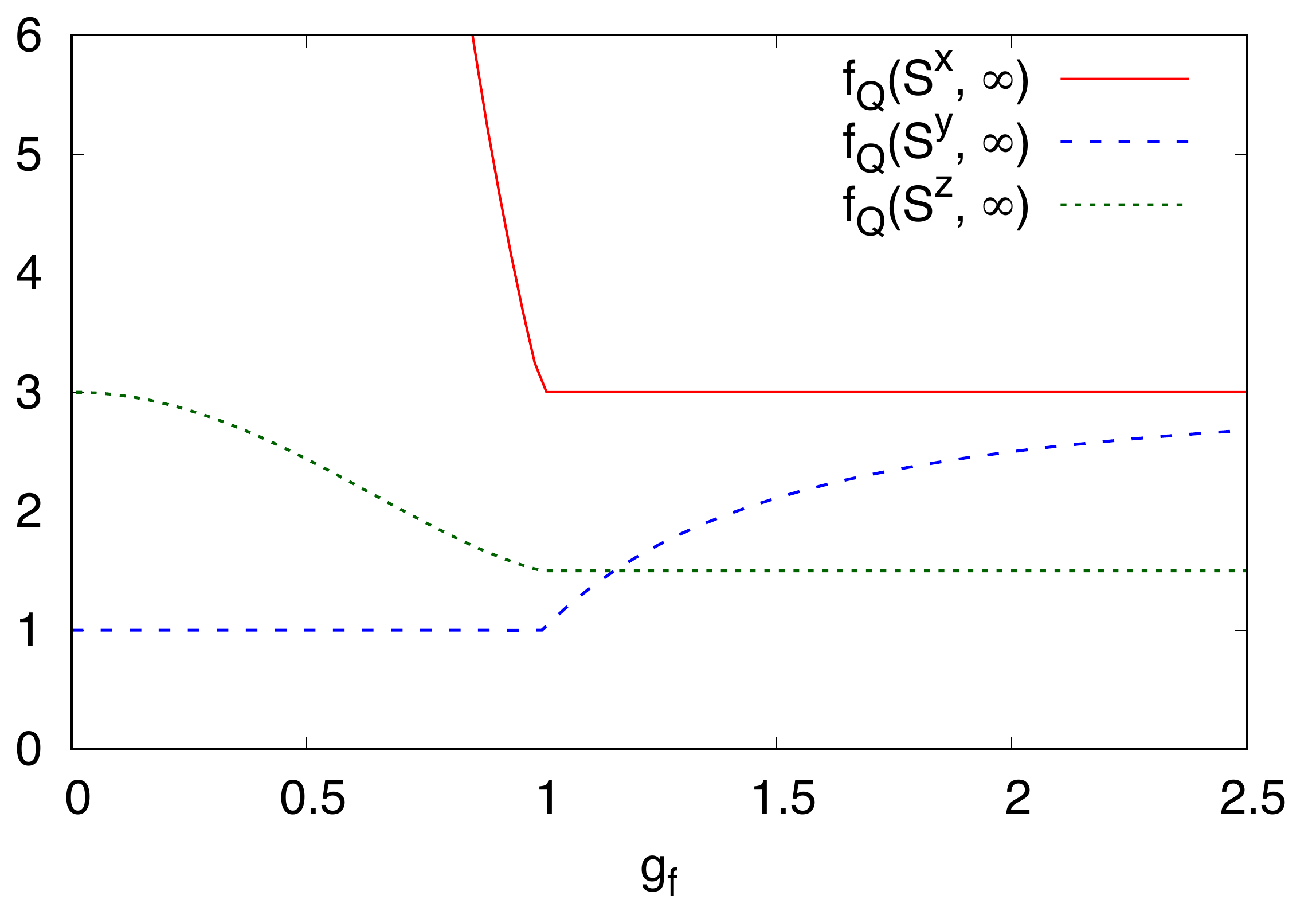}
\centering
 \includegraphics[width=10.5cm]{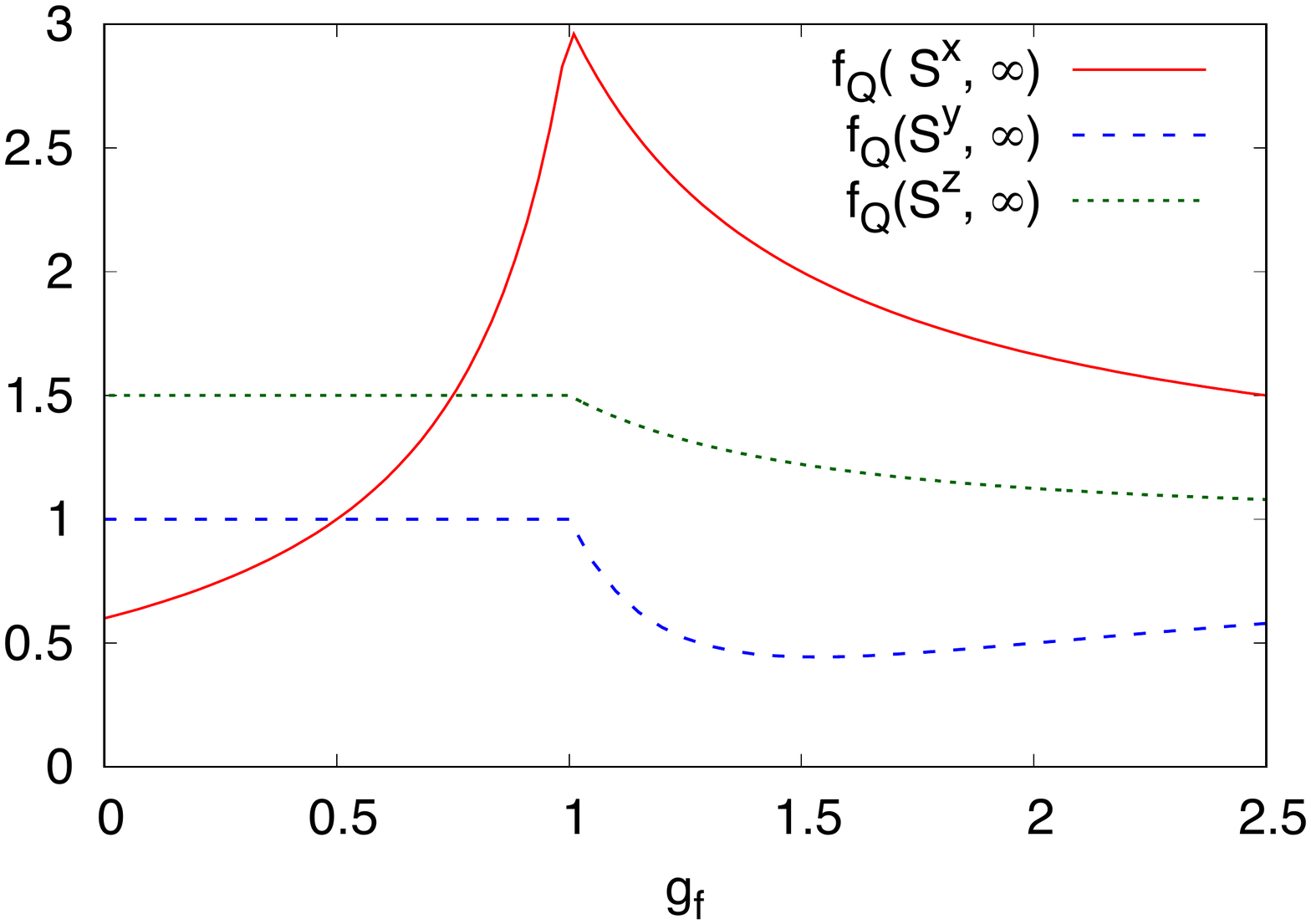}
% \label{fig:minipage1}
\caption{Exact results for the the asymptotic $f_Q$ obtained for the three total spin components $\hat{S}^{x}, \hat{S}^{y} \text{and}\,  \hat{S}^{x}$. The plot shows the behaviour of $f_Q$ as a function of the final field $g_f$ starting from the totally ordered state $g_0=0$ (upper panel) and the fully polarised one $g_0=\infty$ (lower panel). }
\label{fig:compa}
\end{figure}

\begin{equation}
\label{eq:qfi_inf}
f_Q(\hat{S}^x,\infty)=
\begin{cases}
\frac{3}{5-4g_f} \quad \text{for}\,  g_f \leq 1  \, ,\\
\\
\frac{2g_f+1}{2g_f-1} \quad \text{for}\,  g_f \geq 1 \, ;\\
\end{cases}
\end{equation}
while in the opposite limit, when ${g_0=0}$, we find
 
\begin{equation}
\label{eq:qfi_0_dyn}
f_Q(\hat{S}^x,\infty)= 
\begin{cases}
\frac {8-5g_f^2}{g_f^2}  \quad \text{for}\, 0< g_f\leq 1 \, ,\\ \\
3 \quad \text{for} \, g_f \geq 1\, .
\end{cases}
\end{equation}
It is possible to analytically compute also the expression for the QFI relative to the magnetization $\hat{S}^z$, obtaining
for ${g_0=0}$ 
\begin{equation}
\label{eq:tra_g0}
f_Q(\hat{S}^z,\infty)=
\begin{cases}
3 +g_f^4 -\frac 52 g_f^2 \quad \text{for}\,  g_f \leq 1  \, ,\\
\\
\frac 32 \quad \text{for} \, g_f \geq 1\, ;
\end{cases}
\end{equation}
on the opposite side, for ${g_0=\infty}$, the QFI reads
\begin{equation}
\label{eq:tra_ginf}
f_Q(\hat{S}^z,\infty)=
\begin{cases}
\frac 32 \quad \text{for}\,  g_f \leq 1  \\
\\
1 +\frac 1{2\, g_f^2} \quad \text{for} \, g_f \geq 1\,.
\end{cases}
\end{equation}

These functions, together with the corresponding $f_Q(\hat{S}^y,\infty)$ with the same initial conditions, are plotted in Fig.~\ref{fig:compa}.
Information about the degree of multipartiteness of the entanglement can be obtained from the maximum of these three functions, which is 
 $f_Q(\hat{S}^x,\infty)$ in almost all cases, with the exception of $g_f< 3/4$ for $g_0=+\infty$, where $f_Q(\hat{S}^z,\infty)$ dominates. 

On the basis of these data, we can therefore infer that for a maximally ordered initial condition ($g_0=0$), when the final field is larger than the critical value $g_f>1$, all 
the final states display at least tripartite entanglement $f_Q(\infty)=3 \, \forall g_f$. When $0<g_f\leq 1$, the QFI density is greater than three and 
the smaller the quench is, the higher the degree of entanglement. This suggests that the best achievable multipartiteness of 
entanglement is obtained for \emph{infinitesimal} quenches ($g_f=g_0+\epsilon$). We are going to see in the next subsection that this 
is a general feature for $g_0\leq 1$.
For ferromagnetic initial conditions, indeed, there is no limitation on the degree of multipartiteness achievable. 

When the initial condition is fully polarised ($g_0=+\infty$), instead, the structure of entanglement in the steady state is completely different and the multipartiteness
is limites and reaches a maximum when the quench ends at the equilibrium critical point. In particular, while
for $g_f<3/4$ the entanglement is at least bipartite, if we get closer to the critical point the degree
of entanglement grows as well; in particular there is a region ($7/8<g_f<3/2$, Fig.~\ref{fig:compa}) where the entanglement 
is at least tripartite. %{\bf che significa che non \`e genuinamente multipartito?}.
For even larger $g_f$ ($g_f>3/2$) we have $f_Q>1$: in this interval the entanglement finally lapses back to be at least bipartite. 
\subsection{Numerical results for  generic $g_0$}
\label{sec:numer}
Let us now discuss to what extent the features observed above for quenches starting in maximally ordered/disordered states are generic.
In order to do so, let us consider the case of generic $g_0$ and present the results of a numerical evaluation of the QFI.
The asymptotic condition is reached only in the thermodynamic limit, which is numerically attained along the lines described in \ref{app:thermo}. \\

\begin{figure}[ht]
\centering
 \includegraphics[width=12cm]{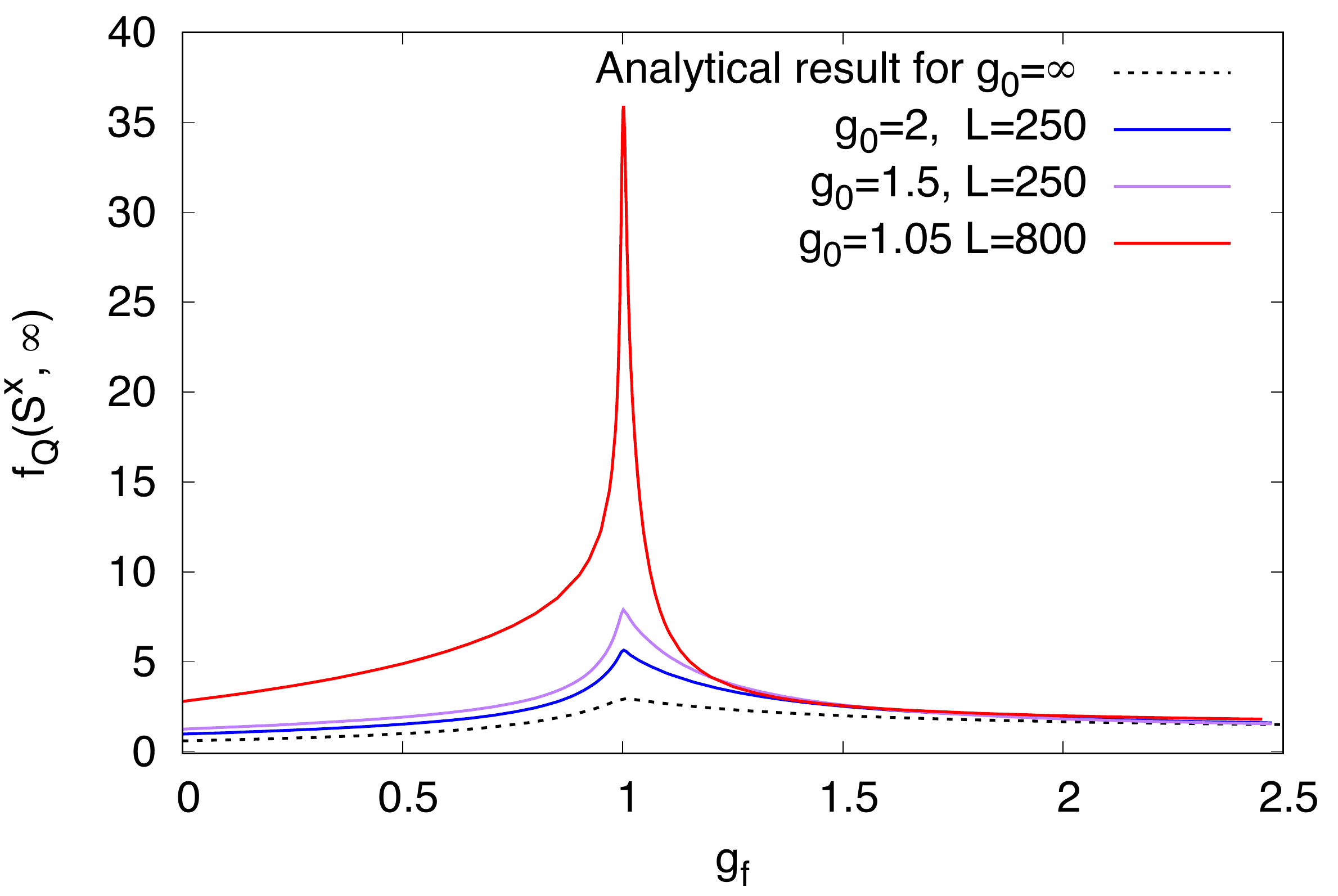}
% \label{fig:minipage1}
%{\bf Angelo: nella label della curva nera c'\`e scritto ``Analitical''. Per favore, Silvia, correggi con ``Analytical''.}
\caption{$\hat{S}^x$ QFI density for the asymptotic state, as a function of the final field. We consider different values of the initial field $g_0$ in the paramagnetic phase (we take also different values of $L$ for issues connected with the convergence to the thermodynamic limit, see~\ref{app:thermo}). 
We take $g_0=2$ (blue plot), $g_0=1.5$ (pink plot) and $g_0=1.05$ (red plot). The numerical results are compared with the exact evaluation 
in the case of $g_0=\infty$, black line in the plot.}
\label{fig:qfi_>1}
\end{figure}

\begin{figure}[htpp]
\centering
 \includegraphics[width=12cm]{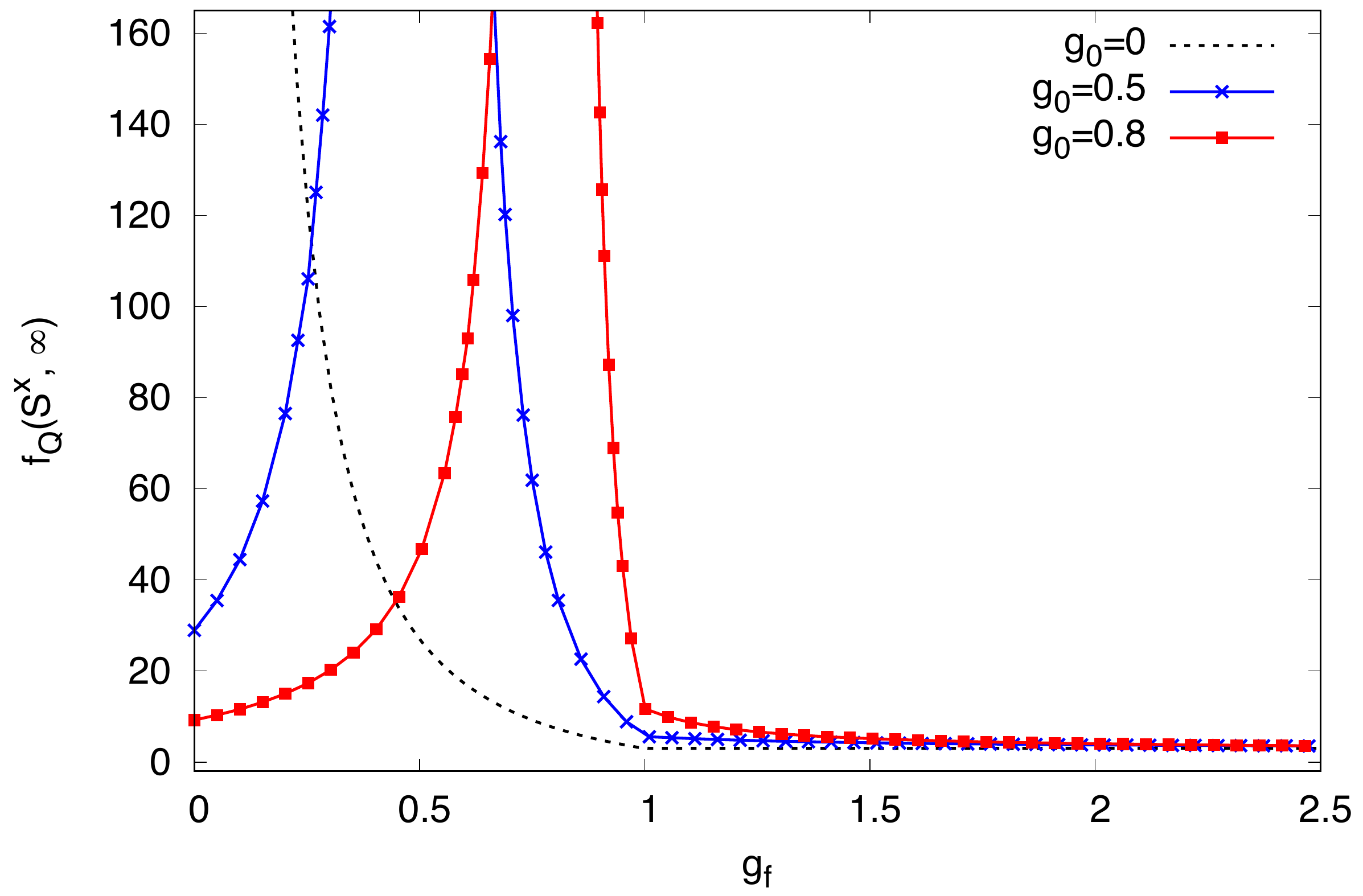}
% \label{fig:minipage1}
\caption{$\hat{S}^x$ QFI density for the asymptotic state, as a function of the final field for different values of the initial field in the ferromagnetic phase. We consider $g_0=0.5$ (blue plot with stars) evaluated for $L=1200$ and 
$g_0=0.8$ (red plot with boxes) evaluated with $L=1200$. The numerical results are compared with the exact evaluation in the case of $g_0=0$ 
(dashed black line in the plot). %For $g_f-g_0>0.2$ we choose $L=1200$, the other points are done with $L=600$.
}
\label{fig:qfi_05}
\end{figure}

Let us start by focusing on $S^x$.
As in the extreme cases described above, the results are significantly different depending on the nature of the initial state.  In particular, whenever 
$g_0>1$, i.e. the initial condition is paramagnetic,
there is a peak at $g_f=1$ which tends to become a divergence in the limit $g_0\to 1$ (see Fig.~\ref{fig:qfi_>1}). 
Therefore, for final $g_f$ close to criticality, the multipartite structure of entanglement is maximised. This is in sharp contrast with the case of 
ferromagnetic initial conditions: here the entanglement is the more multipartite the smaller the quench.
Indeed, we can see a divergence of the QFI density for $g_f\to g_0$ (the so-called \emph{small quench} regime $\delta g \to 0$), which behaves 
as $\sim\delta g^{-2}$ (see Fig.~\ref{fig:qfi_05}).  %,questo dà la divergenza ma si risolve facilmente % $g_f>1$}
This divergence is linked to the behaviour of the correlation length of the spin correlation function as discussed below. 

\begin{figure}[htpp]
\centering
 \includegraphics[width=11cm]{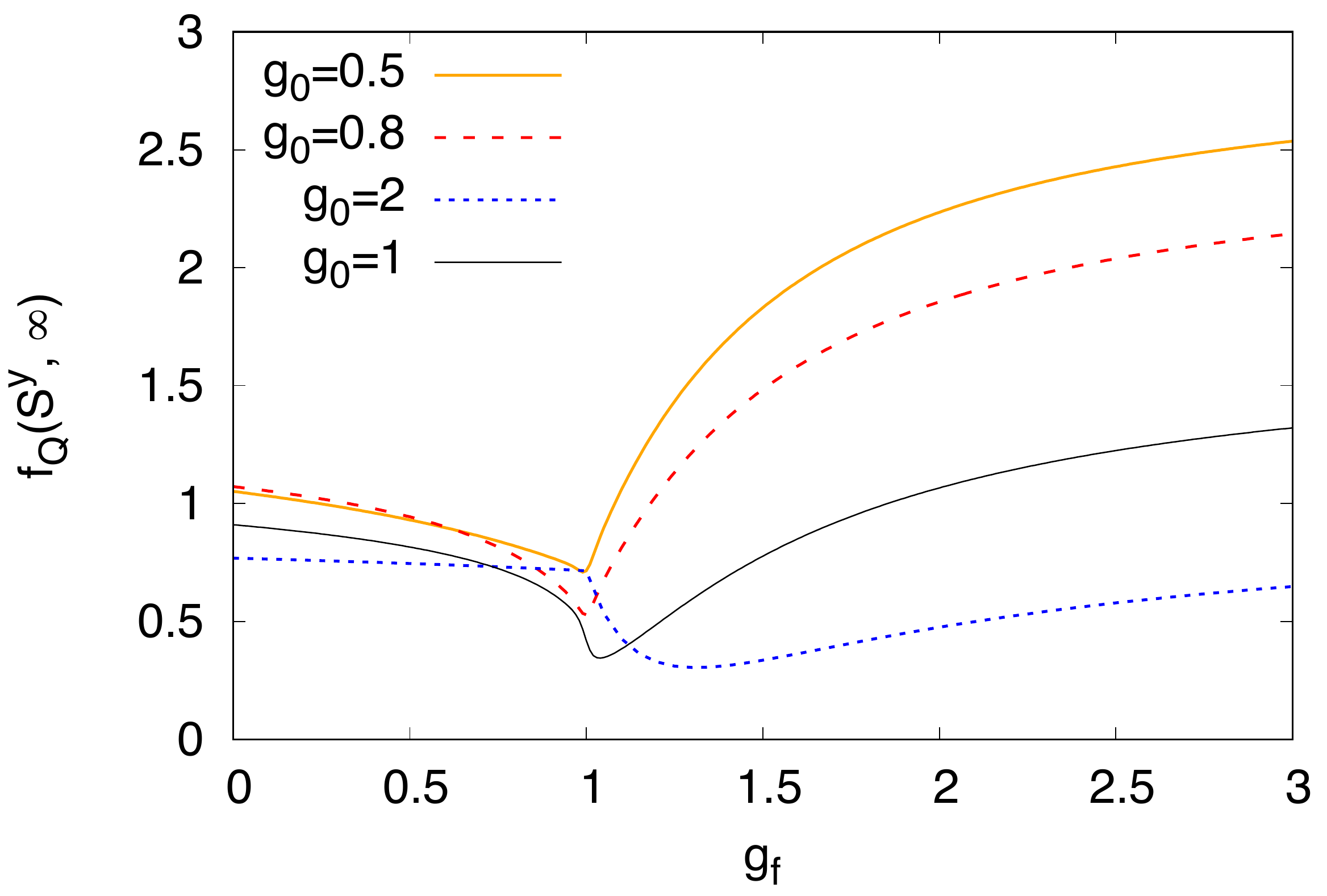}
% \label{fig:minipage1}
\caption{$\hat{S}^y$ QFI density for the asymptotic state to  as a function of the final field $g_f$ for different initial $g_0$ and $L=400$. }
\label{fig:qfi_y_num}
\end{figure}

\begin{figure}[htpp]
\centering
 \includegraphics[width=11cm]{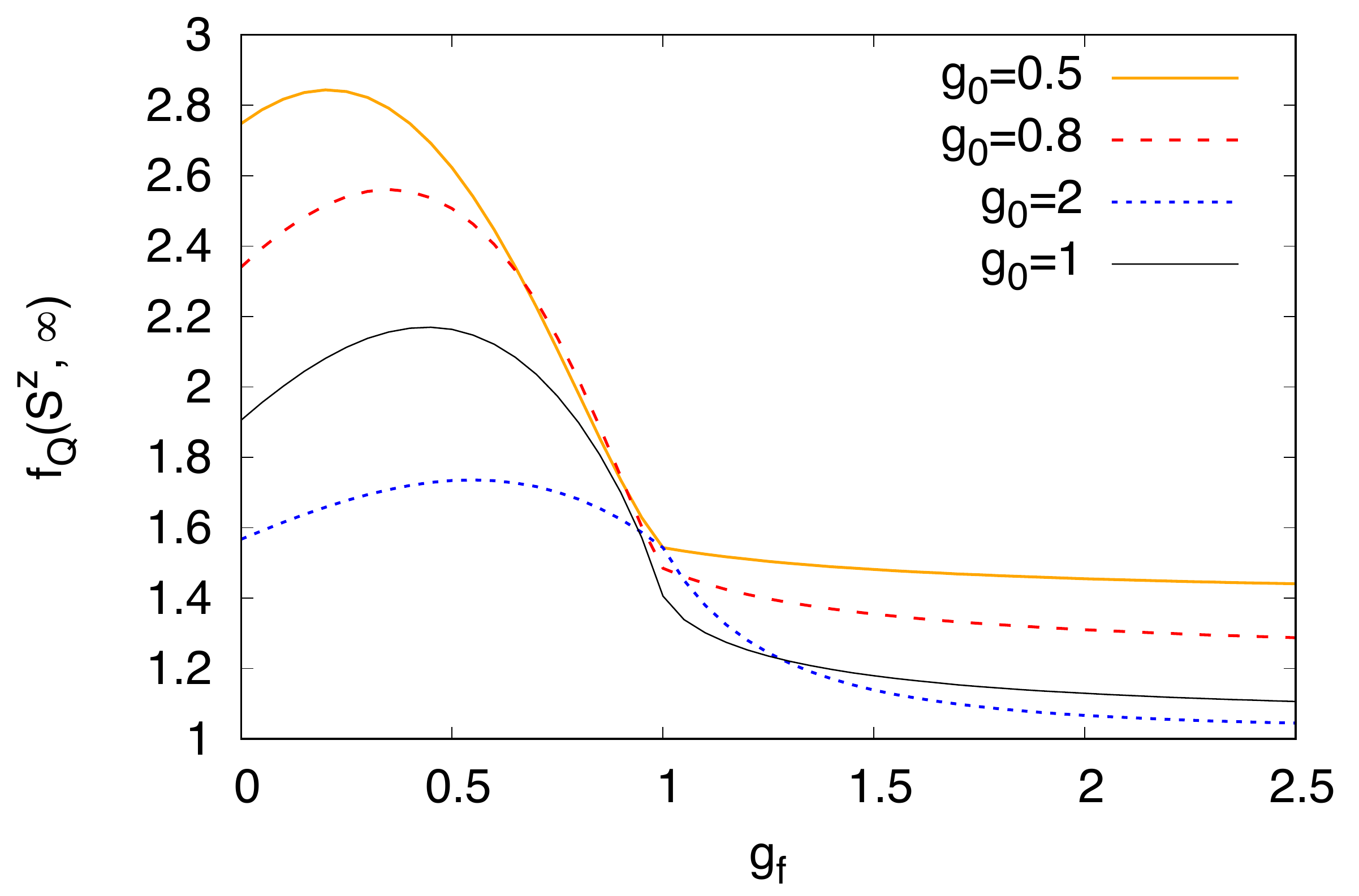}
% \label{fig:minipage1}
\caption{$\hat{S}^z$ QFI density for the asymptotic state as a function of the final field $g_f$ for different initial $g_0$ and $L=400$. 
When $g_f<g*<1$ this gives the optimised QFI density (see Fig. \ref{fig:max_QFI}). }
\label{fig:qfi_z_num}
\end{figure}

Therefore the two main observed features are robust: we see high degree of multipartiteness close to criticality for $0\leq g_0<1$ (system initially in the paramagnetic phase) and
diverging multipartiteness for small quenches when $g_0\geq 1$ (system initially in the ferromagnetic phase). This statement is further corroborated by comparing the graphs for $f_Q(S_x,\infty)$ (Figs.~\ref{fig:qfi_>1} and~\ref{fig:qfi_05})
to those in  Fig. (\ref{fig:qfi_y_num}) and (\ref{fig:qfi_z_num}) where we show the numerical results for the QFI related to to $\hat{S}^y$ and $\hat{S}^z$, for generic $g_0$. We see that, for $g_0>1$, there is an interval of $g_f$ (one of the extrema is 0) where the optimal asymptotic QFI density is given by the $\hat{S}^z$
QFI density, exactly as happens for $g_0=\infty$ (lower panel of Fig.~\ref{fig:compa}).

%
%\paragraph{$g_0<1$} 

%
%Both the height of the peak for $g_0>1$ and the nature of the divergence for $g_0<1$ can be analytically explained
%\paragraph{$g_0>1$, $g_f<1$}
%
%\paragraph{$g_0<1$, $g_f<1$}
%..................................................... Perturbative argument ................................................................................%
\subsection{Perturbative approach to the small-$\delta g$ divergence for quenches with $g_0<1$}
\label{sec:small_quenches}
The divergence for $g_0<1$ can be understood in terms of the corresponding divergence of the correlation length. Consider a ferromagnetic initial
condition: for a whatsoever small quench with $g_f \neq g_0$ the system will not be able to sustain a finite order
parameter in the stationary state, i.e. $\mean{\hat S^x}=0$, while the correlations of the order parameter will exponentially decay
according to $G_n(\infty)\simeq\nep^{-n/\xi}$. 
%When $g_f\to g_0$ the correlation length $\xi$ of the system tends to diverge,
%although $\mean{\hat S^x}$ is always 0. 
Therefore, using Eq.(\ref{eq:qfi_gen_ord}), we can estimate for large $\xi$ the asymptotic QFI as
\begin{equation}
\label{jsjs}
  f_Q(\hat{S}^x,\infty) \sim 1+2\, \xi\,.
\end{equation}
For $g_f\to g_0$, the correlation length diverges, as we can perturbatively find:  following Ref.~\cite{calabrese2011},  the correlator 
behaves as $G_n(\infty)\simeq\nep^{-n/\xi}$ with
\begin{equation}
  \frac{1}{\xi} = - \frac{1}{2\pi}\int_{-\pi}^\pi\ud k \log\left(\frac{g_0g_f-(g_f+g_0)\cos k + 1}{E_k(g_0)E_k(g_f)}\right)\,,
\end{equation}
%
%\sqrt{(1+g_f^2-2g_f\cos k)(1+g_0^2-2g_0\cos k)}
where $E_k(g)=\sqrt{1+g^2-2g\cos(k)}$ is the dispersion of the Ising quasiparticles~\cite{pfeuty1970}. Expanding this expression up to second order in $\delta g$ we find
\begin{equation}
  \frac{1}{\xi} = \delta g^2 \frac{1}{2\pi}\int_{0}^\pi\ud k \, \frac{\sin^2 k}{E_k^4(g_0)}+\mathcal{O}\left(\delta g^3\right) \ .
\end{equation}
Here we can define $\tilde{\xi}(g_0)^{-1}= \frac{1}{2\pi}\int_{0}^\pi\ud k \, \frac{\sin^2 k}{E_k^4(g_0)}= \frac 1{4 (1-g_0)^4}$ for $g_0 \neq 1$. 
This proves therefore that the correlation length for small quenches within the ferromagnetic phase diverges like
\begin{equation}
\label{eq:kkk}
\xi \sim \frac {\tilde{\xi}(g_0)}{\delta g^2}.
\end{equation}
Using now Eq.(\ref{jsjs}) we obtain $ f_Q(\hat{S}^x,\infty) \sim 1/(\delta g)^2$, as expected.
%
%................................................................................................................................................%
%........................................................ Dynamics ........................................................................................%
%''''''''''''''''''''''''''''''''''''''''''''''''''''''''''''''''''''''''''''''''''''''''''''''''''''''''''''''''''''''''''''''''''''''''''''''''%
\subsection{Time dependence of the QFI density}
We conclude this Section briefly commenting on the behaviour of the  QFI density for the order parameter as a function of time.
Here, the correlator is no more given by a Toeplitz determinant as in the asymptotic case  and we need to evaluate a Pfaffian of a more complex
correlation matrix, as elucidated in Ref.~\cite{barouch1971} (from a numerical point of view we use the algorithms and the routines
introduced in Ref.~\cite{pfapack}).

\begin{figure}[htpp]
\centering
 \includegraphics[width=11cm]{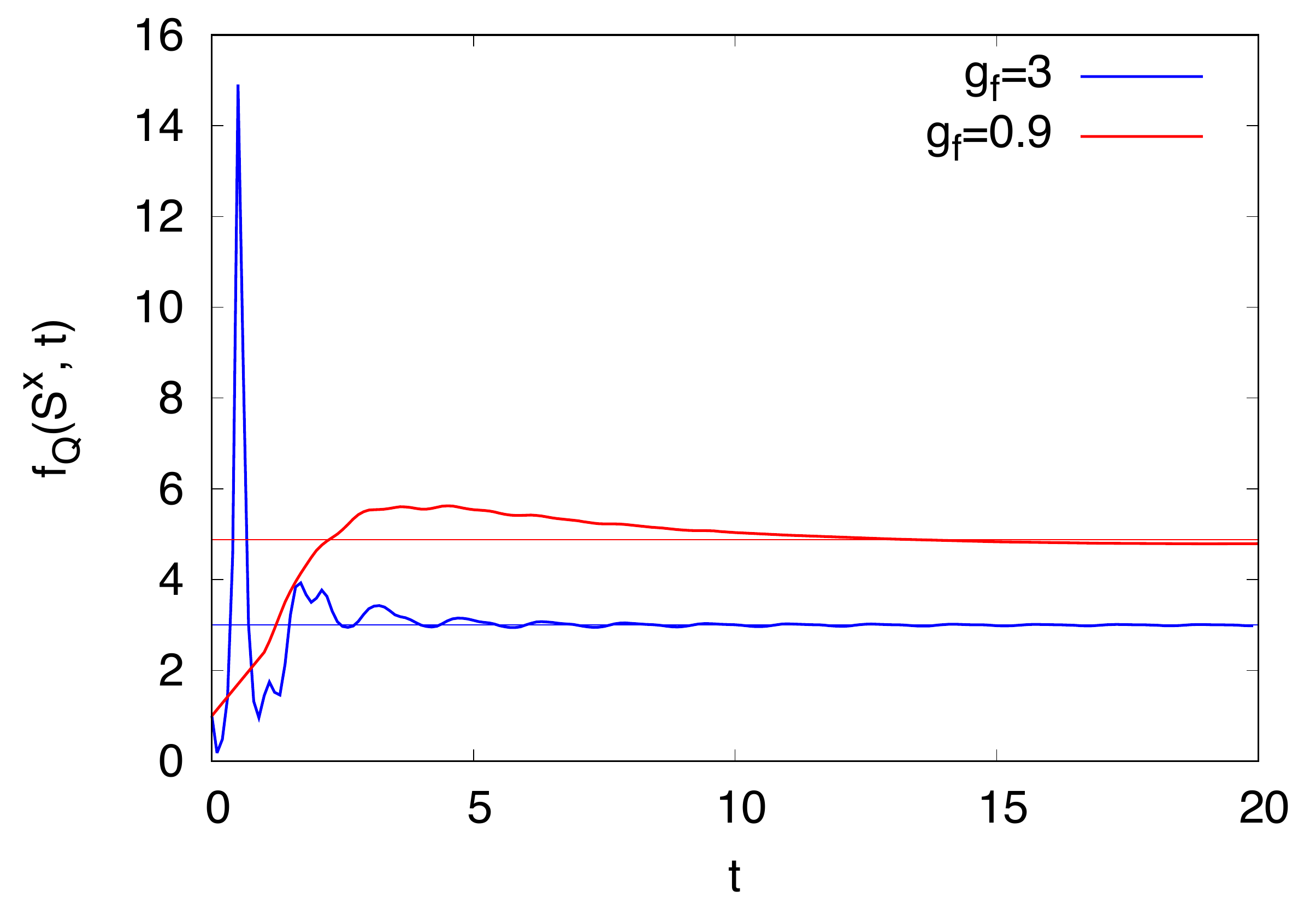}
% \label{fig:minipage1}
\caption{$\hat{S}^x$ QFI density dynamics for a quench from $g_0=0$ to $g_f=0.9$, $g_f=2$. The size of the chain is set to $L=100000$.}
\label{fig:dyn_para}
\end{figure}

\begin{figure}[htpp]
\centering
 \includegraphics[width=11cm]{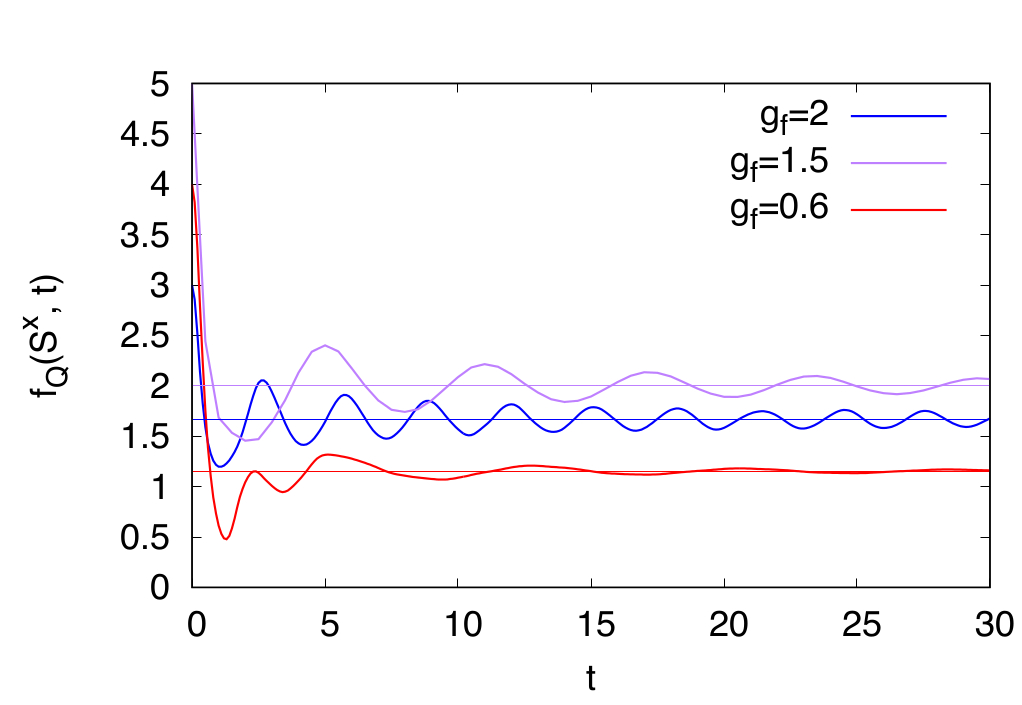}
% \label{fig:minipage1}
\caption{$\hat{S}^x$ QFI density dynamics for a quench from $g_0=\infty$ to $g_f=2$, $g_f=1.5$ and $g_f=0.6$. The size of the chain is set to $L=10^6$. %{\bf Saro: perch\'e non polttiamo fino a $t=30$?} 
}
\label{fig:dyn_ferro}
\end{figure}

From the exact expression of the correlation functions we can also extract how the QFI attains it asymptotic value. Since most of the phase diagram
is dominated by $\hat{S}^x$ let us discuss its Fisher information only  (the other cases are qualitatively very similar).
We report here in Fig.~\ref{fig:dyn_para} some results for $g_0=0$ and in Fig.~\ref{fig:dyn_ferro} some for $g_0=\infty$.
 In the first case we see a peak in the entanglement at short times whose height increases with $g_f$.
 We find that the QFI density tends to the asymptotic value oscillating around it with an amplitude decreasing as a power law.
From the  Fourier transform of the signal it can be clearly seen that the frequency of these oscillations equals the quasi-particle gap~\cite{pfeuty1970} of the final Hamiltonian $\Delta E_0(g_f)=2|1-g_f|$.
%These findings
%occur also for the other considered operators, so -- in order to not bother the reader -- we give examples of evolution only for $\hat{S}^x$. 

%

%as $1/(\delta g)^2$ and so does the QFI density. Remarkably, when $g_0< 1$, this is a singular limit:
% for $g_f=g_0$ we are in the ferromagnetic ground state for which $\mean{\hat S^x}\neq 0$. We have to evaluate the connected correlators (here
% the order parameter is finite). Instead, if we take $g_f-g_0$ infinitesimally small, $\mean{\hat S^x}=0$: the QFI density diverges.

%We can analytically find the divergence of $\xi$ (and then of $f_Q(\hat{S}^x,\infty)$) by means of perturbation theory. 

%This proves thereby that the correlation length for small quenches within the ferromagnetic phase diverges like
%
%................................................................ CONCLUSIONS ..................................................................................%
\section{Conclusions and perspectives} \label{conclusion:sec}
In conclusion, we have studied the multipartite entanglement in a quantum system subjected to a quantum quench. Probing multipartite entanglement through
the Quantum Fisher Information, we have found an expression of the latter in terms of a generalized correlation function.
Considering a quench starting from a pure state and taking the
system in the thermodynamic limit, we have demonstrated that the QFI relaxes to an asymptotic value given by
the fluctuations of an operator in the diagonal ensemble. 
We have then discussed in detail the structure of entanglement in the stationary condition attained by a specific model: the quantum Ising chain 
after a quench in the transverse field. We have found two different scenarios, depending on the initial condition being ferromagnetic or paramagnetic. In the first case,
there is no limitation on the degree of multipartiteness achievable: the smaller the quench the higher the entanglement. In the second,
the degree of multipartiteness is limited (while it tends to diverge as $g_0 \rightarrow 1$) and attains a maximum only close to the equilibrium critical point
$g_f \simeq 1$.
 
A possibility of future work will be to study multipartite entanglement in periodically driven systems, both in the asymptotic condition~\cite{russomanno2012}
and in the Floquet ground state~\cite{russomanno2016,russomanno_arx17} (the latter is an eigenstate of the stroboscopic dynamics which can undergo quantum phase transitions). 
In addition, it would be interesting to address
the quantum Fisher information in disordered systems, both in connection with quantum phase transitions at equilibrium~\cite{fisher1995}
and many body localization~\cite{basko2006}. Especially in the latter case, there are detailed analyses of bipartite entanglement~\cite{mbl-ent,iemini2016,
pietracaprina2016} but an analysis of the multipartite case, with the notable exception of Refs.\cite{goold2015,morone_NP_2016}, is still missing.  
\ack
We thank Pasquale Calabrese, John Goold, Giorgio Parisi, and Boja Zunkovic for useful discussions.
This research was supported in part by UNICREDIT, by the EU integrated projects QUIC and by ``Progetti interni - SNS''. S.P. thanks Accademia Nazionale dei Lincei for financial support under the scholarship ``Enrico Persico''.
\appendix
%
%.......................................... FISHER INFORMATION AND RESPONSE FUNCTIONS .............................................................%
\section{Fisher information and susceptibility out of equilibrium} \label{equilibrium:sec}
We have found that after a quantum quench the QFI can be expressed in terms of a \it generalized \rm response function
of the operator $\cO$ generating the shift, Eq.(\ref{generalFisher}). 
In Ref.~\cite{hauke2016}, considering a quantum many body system at equilibrium, the authors show that QFI can be written as an integral of the linear response susceptibility of the operator $\cO$. 
This is a very important result: it allows to measure the multipartite entanglement in the
laboratory (there are well established experimental methods to measure susceptibilities independently of the size of the considered system). So one would be tempted to directly generalize this result, at least for the stationary state, with a susceptibility averaged over the diagonal ensemble~\cite{polkovnikov2011}. Nonetheless, far from equilibrium, this would be wrong for a general system even for a pure initial state. This follows from the fact that the fluctuation-dissipation theorem~\cite{kubo1986nonequilibrium}, which at equilibrium relates the Keldysh component of the response function to the linear susceptibility as $
\chi''(\omega)= \tanh \left ( \frac{\beta \omega}2 \right ) \chi^K(\omega)
$, is not valid in this general form in the non-equilibrium case \cite{esslerFDT2012}. The susceptibility on the diagonal ensemble, written in Lehmann representation 
\begin{equation}
\bar{\chi}''(\omega)= \pi \sum_{ij}(|c_i|^2 -|c_j|^2) |\mathcal O_{ij}|^2 \delta(\omega-(E_j-E_i))\, , \nonumber
\end{equation}
can not be related in general to the Keldysh component of Eq.(\ref{FQK:eqn}). 
%Exceptions are given by systems whose stationary states are described by thermal distributions with effective temperatures \cite{foiniFDT2012, foiniFDT2011, khatamiFDT2014} {\bf Parlarne con Silvia}. In those cases, a generalization of the result in Ref.~\cite{hauke2016} can be written with an effective temperature. 

%%%%%%%%%%%%%%%%%%%%%%%%%%%%%%%%%%%%%%%%%%%%%%%%%%%%%%%%%%%%%%%%%%%%%%%%%%%%%%%%%%%%%%%%%%%%%%%%%%%%%%%%%%%%%%%%%%%%%%%%%%%%%%%%%%%%%%%%%%%%%%%%%%%%%%%%%%%%%%%%%%%%%
\section{Numerics and the thermodynamic limit}
\label{app:thermo}
\begin{figure}[htp]
\centering
 \includegraphics[width=11cm]{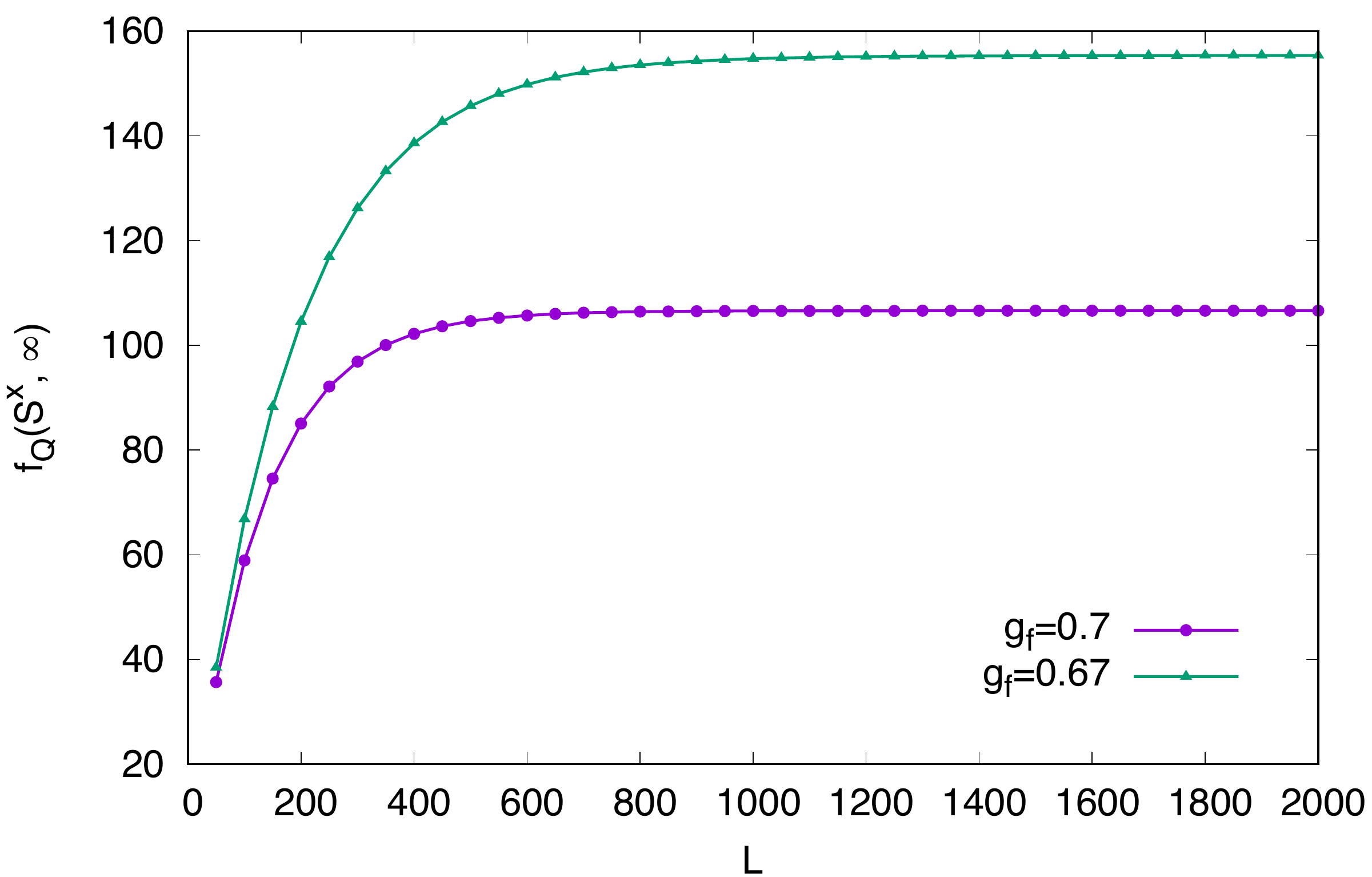}
% \label{fig:minipage1}
\caption{Convergence on the $f_Q(\hat{S}^x, \infty)$ as increasing the length of the chain $L$. We consider quenches within the ferromagnetic phase, where the correlation length diverges with $g_f-g_0\to 0$. We take $g_0=0.5$ and $g_f=0.67$ (purple line) and $g_f=0.7$ (green line).  }
\label{thermo_lim}
\end{figure}

As we have discussed, the stationary state is well defined only in the thermodynamic limit. So, in the numerical evaluation of the asymptotic QFI (see Eq.(\ref{eq:qfi_gen_ord})), we have to be sure that this limit is reached  {\footnote{For the results of Section \ref{sec:exa}, this limit is exact since the correlators are analytic and the series converges.}}. For each $g_0$ and $g_f$, we choose $L$ in order to satisfy $L \gg \xi(g_0, g_f)$. When this condition is met, the thermodynamic limit is reached and the asymptotic QFI is well defined. 

\noindent As an example, in Fig.~\ref{thermo_lim} we show the convergence of $f_Q(S^x, \infty)$ to its well-defined thermodynamic limit value, by increasing the length of the chain $L$. The value becomes constant above a certain value of $L$. Notice that in the ferromagnetic phase, since $\xi \sim \delta g ^{-2}$, for smaller $\delta g$ one should consider longer chains.


\section*{Bibliography}
\vspace{0.5cm}

\begin{thebibliography}{99}

\bibitem{amico2008}
	L. Amico, R. Fazio, A. Osterloh, and V. Vedral, Rev. Mod. Phys. {\bf 80}, 517 (2008).
\bibitem{calabrese2009}
	P. Calabrese, J. Cardy, and B. Doyon Eds, J. Phys. A {\bf 42}, 500301 (2009).
\bibitem{eisert2010}
	J. Eisert and M. Cramer and M.B. Plenio, Rev. Mod. Phys. {\bf 82}, 227 (2010).
\bibitem{kitaev2006}
	A. Kitaev and J. Preskill, Phys. Rev. Lett. {\bf 96}, 110404 (2006).
\bibitem{li2008}
	H. Li and F. D. M. Haldane, Phys. Rev. Lett. {\bf 101}, 010504 (2008).
\bibitem{osterloh2002}
	A. Osterloh, L. Amico, G. Falci, and R. Fazio,  Nature {\bf 416}, 608 (2002).
\bibitem{vidal2003}
	G. Vidal, J.I. Latorre J.I. , E. Rico, and A. Kitaev, Phys. Rev. Lett. {\bf 90}, 227902 (2003).
\bibitem{calabrese2004}
	P. Calabrese and J. Cardy, J. Stat. Mech. P06002, (2004).
\bibitem{polkovnikov2011}
	A. Polkovinkov, K. Sengupta, A. Silva, and M. Vengalattore, Rev. Mod. Phys. {\bf 83}, 863 (2011).
\bibitem{calabrese2005}
	P. Calabrese and J. Cardy, J. Stat. Mech. P04010, (2005).
\bibitem{dechiara2006}
	G. De Chiara, S. Montangero, P. Calabrese, and R. Fazio,  J. Stat. Mech. P03001 (2006).
\bibitem{mbl-ent} 
	M. Znidaric, T. Prosen, and P. Prelovsek, Phys. Rev. B {\bf 77}, 64426 (2008).
\bibitem{mbl-ent1} 
	J. H. Bardarson, F. Pollmann, and J. E. Moore, Phys. Rev. Lett. {\bf 109}, 017202 (2012).
\bibitem{mbl-ent2} 
	M. Serbyn, Z. Papic, and Dmitry A. Abanin, Phys. Rev. Lett. {\bf 110}, 260601 (2013).
\bibitem{kz-ent}
	R. W. Cherng and L. S. Levitov, Phys. Rev. A, {\bf 73}, 043614 (2006).
\bibitem{kz-ent1}
	L. Cincio, J. Dziarmaga, M. M. Rams, and W. H. Zurek,  Phys. Rev. A, {\bf 75}, 052321 (2007).
\bibitem{kz-ent2}
	T. Caneva, R. Fazio, and G.E. Santoro, Phys. Rev. B, {\bf 78} 104426 (2008).
\bibitem{pd-ent}
	A. Sen, S. Nandy,  and K. Sengupta,   Phys. Rev. B {\bf 94}, 214301 (2016).
\bibitem{pd-ent1}
	V. Eisler and I. Peschel, Annalen der Physik, {\bf 17}, 410 (2008).
\bibitem{pd-ent2}
	P. Barmettler, A.-M. Rey, E. Demler, M.D. Lukin, I.l Bloch, and V. Gritsev, Phys. Rev. A, {\bf 78}, 012330 (2008).
\bibitem{pd-ent3}
	A. Russomanno, G.E. Santoro, and R. Fazio, J. Stat. Mech. 073101 (2016).
\bibitem{pd-ent4}
	T. J. G. Apollaro, G. M. Palma, and J. Marino, Phys. Rev. B {\bf 94}, 134304 (2016).
\bibitem{toth2012}
	G. T\`oth, Phys. Rev. A 85, 022322 (2012).
\bibitem{boileau2004}	
	J.-C. Boileau, D. Gottesman, R. Laflamme, D. Poulin, and R. W. Spekkens, Phys. Rev. Lett. {\bf 92}, 017901 (2004).
\bibitem{oliveira2006}
	T. R. de Oliveira, G. Rigolin, M. C. de Oliveira, and E. Miranda, Phys. Rev. Lett. {\bf 97}, 170401 (2006).
\bibitem{wei2005}	
	T.-C. Wei, D. Das, S. Mukhopadyay, S. Vishveshwara, and P. M. Goldbart, Phys. Rev. A {\bf 71}, 060305 (2005).
\bibitem{biswas2014}
	A. Biswas, R. Prabhu, A. Sen(De), and U. Sen, Phys. Rev. A {\bf 90}, 032301 (2014).
\bibitem{hofmann2014} 
	M. Hofmann, A. Osterloh, and O. G\"uhne, Phys. Rev. B {\bf 89}, 134101 (2014).
\bibitem{gulden2016}	
	T. Gulden, M. Janas, Y. Wang, and A. Kamenev, Phys Rev. Lett. {\bf 116}, 026402 (2016).
\bibitem{girolami1}
        D. Girolami and B. Yadin, Entropy {\bf 19} (3), 124, (2017).
\bibitem{wong2001}
	 A. Wong and N. Christensen, Phys. Rev. A {\bf 63}, 044301 (2001).
%
\bibitem{Nielsen_Chuang:book}
M.~Nielsen and I.~L. Chuang.
\newblock {\em Quantum Computation and Quantum Information}.
\newblock Cambridge University Press, 2000.
%
\bibitem{meyer2002}	 
	 D. A. Meyer and N. R. Wallach, J. Math. Phys. {\bf  43}, 1497700 (2002).
\bibitem{osterloh2005}
	A. Osterloh and J. Siewert, Phys. Rev. A {\bf 72}, 012337 (2005).
\bibitem{oliveira_pra_2006}	 
	 T. R. de Oliveira, G. Rigolin, and M. C. de Oliveira, Phys. Rev. A {\bf 73}, 010305 (2006).
\bibitem{facchi2010}
	 P. Facchi, G. Florio, U. Marzolino, G. Parisi, S. Pascazio, J. Phys. A: Math. Theor. 43, 225303 (2010).
 \bibitem{facchi2008maximally}
 P. Facchi, G. Florio, G. Parisi and S. Pascazio, Physical Review A, {\bf 77}, 060304 (2008).
\bibitem{guhne2009}
	O. G\"uhne and G. T\`oth, Phys. Rep. {\bf 474}, 1 (2009).
\bibitem{guhne2005}
	O. G\"uhne, G. T\`oth, and H. Briegel, New J. Phys. {\bf 7}, 229 (2005).
\bibitem{strobel2014}	
	H. Strobel, W. M\"ussel, D. Linnemann, T. Zibold, D.B. Hume, L. Pezz\`e, A. Smerzi, and M. Oberthaler, Science {\bf 345}, 424 (2014).
\bibitem{hyllus2012}
	P. Hyllus, W. Laskowski, R. Krischek, C. Schwemmer, W. Wieczorek, H. Weinfurter, L. Pezz\`e, and A. Smerzi,
	Phys. Rev. A {\bf 85}, 022321 (2012).
\bibitem{braunstein94} S.~L. Braunstein and C.~M. Caves, Phys. Rev. Lett. {\bf 72}, 3439 (1994).
\bibitem{hauke2016}
	P. Hauke,	M. Heyl, L. Tagliacozzo, and P. Zoller, Nat. Phys. {\bf 12}, 778 (2016). 
	%
\bibitem{toth2014quantum}
G. T{\'o}th, and I. Apellaniz, Journal of Physics A: Mathematical and Theoretical {\bf 42}, 424006 (2014).
%
\bibitem{apellaniz2015optimal}
I. Apellaniz, M. Kleinmann, O. G{\"u}hne and G. T{\'o}th,
arXiv preprint arXiv:1511.05203, (2015).
%
\bibitem{Lieb_AP61}
E.~Lieb, T.~Schultz, and D.~Mattis,
%\newblock Two soluble models of an antiferromagnetic chain.
\newblock {\em Annals of Physics}, 16:407--466, 1961.

%\bibitem{Pfeuty_AP70}
%Pierre Pfeuty.
%\newblock The one-dimensional Ising model with a transverse field.
%\newblock {\em Annals of Physics}, 57:79--90, 1970.
%
\bibitem{horodecki2009}
	R. Horodecki, P. Horodecki, M. Horodecki, and K. Horodecki, Rev. Mod. Phys. {\bf 81}, 865 (2009). 
\bibitem{sen2010}	
	A. Sen(De) and U. Sen, Phys. Rev. A 81, 012308 (2010).
\bibitem{holevo-book}
	A. Holevo, {\em Probabilistic and statistical aspects of quantum theory}, Monographs (Scuola Normale Superiore),
	(Springer Science $\&$ Business Media, 2011).
%\bibitem{g_K_o_book}
%        C. Gerry and P. Knight, {\em Introductory Quantum Optics}, Cambridge (2004).
\bibitem{braunstein1994}	
	S. Braunstein and C. Caves, Phys. Rev. Lett. {\bf 72}, 3439 (1994).
%\bibitem{kamenev-book}
%	A. Kamenev, {\em Field Theory of non-equilibrium systems}, Cambridge University Press (Cambridge, 2011).
\bibitem{ziraldo2012}	
	S. Ziraldo, A. Silva, and G. E. Santoro, Phys. Rev. Lett. {\bf 109}, 247205 (2012).
\bibitem{ziraldo2013}		
	S. Ziraldo and G. E. Santoro, Phys. Rev. B {\bf 87}, 064201 (2013).	
\bibitem{russomanno2012}			
	A. Russomanno, A. Silva, and G. E. Santoro, Phys. Rev. Lett {\bf 109}, 257201 (2012)
\bibitem{pfeuty1970}
	P. Pfeuty, Ann. Phys. {\bf 57}, 79 (1970).
%\bibitem{note2}	
%	Taking anti-periodic boundary conditions corresponds to restricting to the subsector of the Hilbert space with an even 
%	number of fermions. This is the appropriate choice because this is the sector where there are the initial ground state 
%	$\left |\psi _{\protect \rm  GS}\right \rangle$ {for a chain of finite length $L$} and the ensuing time-evolved state 
%	$\left |\psi (t)\right \rangle$ live [G. Kells, D. Sen, J. K. Slingerland, and S. Vishveshwara,
%	Phys. Rev. B {\bf 89}, 235130 (2014)].
\bibitem{calabrese2011}
	P. Calabrese, F. Essler, and M. Fagotti, Phys. Rev. Lett. {\bf 106}, 227203 (2011).
%
\bibitem{caneva2011}	
	T. Caneva, E. Canovi, D. Rossini, G. E. Santoro, and A. Silva, J. Stat. Mech. , P07015 (2011).	
%
%\bibitem{Caneva_JSM11}
%Tommaso Caneva, Elena Canovi, Davide Rossini, Giuseppe~E. Santoro, and
%  Alessandro Silva,
%
%\newblock Applicability of the generalized gibbs ensemble after a quench in the
%  quantum ising chain.
\newblock {\em J. Stat. Mech.}, page P07015, 2011.
\bibitem{gge}
	E. T. Jaynes, Phys. Rev. {\bf 108}, 171 (1957).
\bibitem{gge4} M. Eckstein and M. Kollar, Phys. Rev. Lett. {\bf 100}, 120404 (2008).
\bibitem{gge1}
	A. Iucci and M. A. Cazalilla, PRA {\bf 80}, 063619 (2009).
\bibitem{gge3} S. R. Manmana, S. Wessel, R. M. Noack, and A. Muramatsu,
	Phys. Rev. Lett. {\bf 98}, 210405 (2007).
\bibitem{gge2}  M. Rigol, V. Dunjko, V. Yurovsky, and M. Olshanii, Phys. Rev. Lett. {\bf 98}, 050405 (2007).
%
\bibitem{gge5} M. Kollar and M. Eckstein, Phys. Rev. A {\bf 78}, 013626 (2008).
%
\bibitem{barouch1971}
	E. Barouch and B. M. M. Coy, Phys. Rev. A {\bf 3}, 786 (1971).
%
\bibitem{pfapack}
M.~Wimmer,
\newblock ACM Trans. Math. Software {\bf 38}, 30 (2012),
\newblock arXiv:1102.3440.
%
\bibitem{sengupta2004}
	K. Sengupta, S. Powell, and S. Sachdev, Phys. Rev. A {\bf 69}, 053616 (2004).
\bibitem{suzuki-book}
	S. Suzuki, J.-I. Inoue, and B. K. Chakrabarti, {\it Quantum Ising phases and transitions in transverse Ising models},  
	Lecture Notes in Physics Vol. 862 (Springer-Verlag Berlin Heidelberg, 2012).
\bibitem{russomanno2016}
	A. Russomanno and E.~G. Dalla Torre, EPL {\bf 115}, 30006 (2016).
\bibitem{russomanno_arx17}
        A. Russomanno, B. Friedman, and E.~G. Dalla Torre, arXiv:1611.00659
\bibitem{fisher1995}
	D. S. Fisher, Phys. Rev. B {\bf 51}, 6411 (1995).
\bibitem{basko2006}	
	D. M. Basko, I. L. Aleiner, and B. Altshuler, Ann. Phys. {\bf 321}, 1126 (2006).
\bibitem{iemini2016}
	F. Iemini, A. Russomanno, D. Rossini, A. Scardicchio, and R. Fazio, (2016), Phys. Rev. {\bf B} 94, 214206.
\bibitem{pietracaprina2016}	
	F. Pietracaprina, G. Parisi, A. Mariano, S. Pascazio, and A. Scardicchio, arXiv:1610.09316.
\bibitem{goold2015}
	J. Goold, C. Gogolin, S. R. Clark, J. Eisert, A. Scardicchio, A. Silva, Phys. Rev. B {\bf 92}, 180202(R) (2015).
%
\bibitem{morone_NP_2016} J. Smith, A. Lee, P. Richerme, B. Neyenhuis, P.~W. Hess, P. Hauke, M. Heyl, D.~A. Huse, and C. Monroe, 
Nature Physics {\bf 12}, 907-911 (2016) .
%
\bibitem{kubo1986nonequilibrium}
R. Kubo, M. Toda, and N. Hashitsume, {\it Non Equilibrium Statistical Mechanics, Statistical Physics II}, Berlin: Springer (1986).
%
\bibitem{esslerFDT2012}
F.~H.~L. Essler and S. Evangelisti, and M. Fagotti, Phys. Rev. Lett. {\bf 109}, 247206 (2012).
%\bibitem{foiniFDT2011}
%Foini, L. and Cugliandolo, L.F. and Gambassi, A., Phys. Rev. B {\bf 84}, 212404 (2011).
%\bibitem{foiniFDT2012}
% Foini, L. and Cugliandolo, L.F. and Gambassi, A., Journal of Statistical Mechanics: Theory and Experiment {\bf 2012}, 09011 (2012).
%
% \bibitem{khatamiFDT2014}
% E. Khatami, G. Pupillo, M. Srednicki and M. Rigol, Journal of Physics: Conference Series, {\bf 510}, 012035 (2014).
% \bibitem{facchi2008maximally}
% Facchi, P- and Florio, G. and Parisi, G. and Pascazio, S., Physical Review A, {\bf 77}, 060304 (2008).
\end{thebibliography}
\end{document}